\documentclass[aps,prx,twocolumn,superscriptaddress,longbibliography,nofootinbib]{revtex4-2}
\usepackage{amsmath,amssymb,bm,mathtools}
\usepackage{graphicx}
\usepackage[colorlinks=true,linkcolor=blue,citecolor=blue,urlcolor=blue]{hyperref}
\usepackage{libertine}

\newcommand{\zhat}{\hat{\bm z}}
\newcommand{\qv}{\bm q}
\newcommand{\Bv}{\bm B}
\newcommand{\bv}{\bm b}
\newcommand{\kv}{\bm k}
\newcommand{\dv}{\bm d}
\newcommand{\gv}{\bm\gamma}
\newcommand{\etav}{\bm\eta}
\newcommand{\so}{\mathrm{so}}
\newcommand{\eSDE}{\eta_{\mathrm{SDE}}}
\newcommand{\Tcs}{T_{\mathrm{cs}}}
\newcommand{\Tct}{T_{\mathrm{ct}}}
\newcommand{\Imm}{\operatorname{Im}}
\newcommand{\Ree}{\operatorname{Re}}
\newcommand{\Aeff}{\mathcal{A}}
\newcommand{\Beff}{\mathcal{B}}

\begin{document}

\title{Superconducting diode effect from field-induced $s+if$ pairing in
Ising superconductors}

\author{Alex Levchenko}
\affiliation{Department of Physics, University of Wisconsin-Madison, Madison, Wisconsin 53706, USA}

\author{Maxim Khodas}
\affiliation{Racah Institute of Physics, Hebrew University of Jerusalem, Jerusalem 91904, Israel}

\date{September 1, 2026}

\begin{abstract}
The in-plane critical field of Ising superconductors exceeds the Pauli
limit by an order of magnitude because the Ising spin-orbit coupling locks
the electron spins out of the basal plane. The same locking converts an
in-plane Zeeman field into a source of equal-spin triplet Cooper pairs, so
that the field-driven condensate acquires an $s+if$ character. We show
that this conversion channel also generates Lifshitz invariants, the
odd-in-momentum terms of the Ginzburg-Landau expansion that are
responsible for the superconducting diode effect. When the basal mirror
symmetry of the monolayer is lifted by a substrate or a gate, the
field-induced triplets couple linearly to the Cooper-pair momentum, and an
intrinsic diode response develops whose strength is set by the
deformability of the band spin texture, quantified by the ratio of the
Zeeman and spin-orbit energies, rather than by the small ratio of the
spin-orbit and Fermi energies familiar from parity-mixing mechanisms. We
construct the symmetry-constrained two-component Ginzburg-Landau theory of
the coupled singlet and triplet order parameters and derive all of its
coefficients from the microscopic model of an Ising superconductor,
finding that the complete functional, including all gradient and quartic
terms, is generated by a single pair-breaking function of temperature,
field, and Cooper-pair momentum. An attractive triplet channel reshapes
the diode response in a characteristic way: at weak fields it suppresses
the efficiency through destructive interference between the direct and the
collective-mode conversion paths, while at strong fields it extends the
diode regime well beyond the singlet-only critical field, with the maximal
efficiency reached along the triplet-enhanced phase boundary. The diode
effect thereby serves as a transport probe of a hidden triplet pairing
channel and of the field-induced $s+if$ state.
\end{abstract}

\maketitle

%=====================================================================
%=====================================================================
%=====================================================================
%=====================================================================
%=====================================================================
\section{Introduction}
\label{sec:intro}

The superconducting diode effect (SDE), a difference between the critical
currents that a superconductor can sustain in opposite directions, and its
device counterpart, the Josephson diode effect, remain active topics of
current research \cite{Ando2020,Nadeem2023,Jiang2022,Ma2025,
ShafferLevchenko2025}. As a bulk property of a material, the SDE is of
fundamental interest: diodicity in a device can be engineered through the
junction geometry, asymmetric barriers, or trapped flux
\cite{Baumgartner2022,Wu2022,Lyu2021}, and does not necessarily reflect
intrinsic material properties. Even in the bulk, extrinsic mechanisms play
an important role, including asymmetric surface and edge barriers,
Meissner screening, and vortex ratchet physics
\cite{Vodolazov2005,Hou2023,Sundaresh2023,Gutfreund2023}; the intrinsic
effect, rooted in the structure of the condensate itself, has to be
separated from these backgrounds. The theory of the intrinsic SDE has been
developed most completely for two classes of two-dimensional systems with
strong spin-orbit coupling: Rashba superconductors
\cite{Edelstein1996,Daido2022,Yuan2022,He2022,Ilic2022,Hasan2024,Hasan2025} and,
more recently, Ising superconductors \cite{Bankier2025}, and it has been
reviewed in Refs.~\cite{Nadeem2023,Jiang2022,Ma2025,ShafferLevchenko2025}.

The symmetry conditions for the SDE are well understood: both inversion
$\mathcal P$ and time reversal $\mathcal T$ must be broken
\cite{Wakatsuki2018,Hoshino2018,Zinkl2022,Daido2022b}. In the
Ginzburg-Landau (GL) description the effect arises from terms of the free
energy that are odd in the collective Cooper-pair momentum $\qv$, commonly
known as Lifshitz invariants
\cite{Mineev1994,Edelstein1996,Agterberg2003,Kaur2005,Dimitrova2007,
MineevSamokhin2008,Kochan2023}. In the standard situation $\mathcal T$ is
broken by a magnetic field $\Bv$. Because $\Bv$ is an axial vector, even
under inversion, the Lifshitz invariants require the crystal itself to
break $\mathcal P$; they exist only in noncentrosymmetric materials, and
their form for the different crystallographic point groups has been
tabulated \cite{Agterberg2012}.

The crystallographic symmetry imposes constraints that go beyond the
absence of an inversion center. Consider a two-dimensional monolayer lying
in the $xy$ basal plane and subject to an in-plane magnetic field. Ising
superconductors such as monolayer NbSe$_2$, TaS$_2$, and gated MoS$_2$
\cite{Xi2016,Costanzo2016,Lu2015,Saito2016,delaBarrera2018,Sohn2018,Dvir2018} are
defined by the basal mirror symmetry $\sigma_h$: monolayers with prismatic
coordination have the point group of a right triangular prism, $D_{3h}$.
The Ising spin-orbit coupling (SOC) polarizes the Bloch states out of
plane with a splitting $\Delta_\so$ that typically exceeds the
superconducting gap by orders of magnitude, which underlies the large
critical fields \cite{Bulaevskii1976,Frigeri2004,Ilic2017,Sosenko2017,
Mockli2018}. An in-plane field breaks $\sigma_h$ and $\mathcal T$
separately but preserves their product $\sigma_h\mathcal T$. This combined
symmetry is instrumental both in the robustness of Ising
superconductivity and in the conversion of singlet Cooper pairs into
triplets mediated by the joint action of the field and the Ising SOC
\cite{Mockli2019,Mockli2020,Haim2020,Tang2021}. The Lifshitz invariant is
odd under $\sigma_h\mathcal T$ and is therefore forbidden in $D_{3h}$ for
any in-plane field, even though the group lacks an inversion center. In
two-dimensional systems without the basal mirror the lowest allowed
invariant reads $\propto\zhat\cdot(\Bv\times\qv)$ with $\qv$ in the basal
plane; the prefactor depends on the band-structure details and on the
strength of the spin-orbit interaction. The point groups consistent with
the broken basal mirror are those of a regular triangular pyramid,
$C_{3v}$, or a cone, $C_{\infty v}$, and the same reduction of symmetry
gives rise to a generalized Rashba spin-orbit splitting of the electronic
bands \cite{BychkovRashba1984,GorkovRashba2001}. In practice the
reduction $D_{3h}\to C_{3v}$ is realized by a substrate or by gating.
Even though the weak Rashba SOC has a negligible effect on the phase diagram \cite{Harms2026},
it can serve as a probe of unconventional Cooper pairing \cite{Hanis2024}.
Moreover, the interplay of the dominant Ising and Rashba SOC has been
shown to strongly enhance the diode response compared to a pure Rashba
superconductor \cite{Bankier2025}.

In this work we identify a mechanism by which the Lifshitz invariants, and
with them the intrinsic SDE, are generated by field-induced triplet Cooper
pairs. Spin-triplet correlations induced by spin-orbit coupling were first
discussed for the Rashba superconductor
\cite{GorkovRashba2001,BauerSigrist2012,Smidman2017}: the admixture of the
triplet component to the Cooper-pair wave function is there a small
fraction, of order $(\Delta_\so/E_F)^2\ll1$, and the same conclusion holds
for a generic form of the spin-orbit interaction, including the Ising one.
In Refs. \cite{Mockli2019,Mockli2020} it shown that triplet correlations of a different
type appear in the presence of spin-orbit coupling. These triplets are generated because the
spin texture of the spin-polarized bands is smoothly deformed by an
applied magnetic field; since the rigidity of the texture is set by the
spin-orbit coupling, the amplitude of the field-induced triplets scales as
$\Delta_Z/\Delta_\so$, where $\Delta_Z$ is the Zeeman energy. This ratio
is insensitive to the large Fermi-energy scale. In an Ising
superconductor with an attractive triplet channel the induced pairs
condense into a coherent $s+if$ state \cite{Mockli2019}, with
thermodynamic and spectroscopic signatures analyzed in
Refs.~\cite{Mockli2020,Haim2020,Tang2021}, consistent with the 
recent tunneling spectroscopy measurements \cite{Kuzmanovic2022}.
Here we explore this conversion
channel in the context of the SDE. We show that once the basal mirror is
lifted, the field-induced triplet order parameter couples linearly to the
Cooper-pair momentum, so that its elimination from the free energy
produces Lifshitz invariants and the accompanying higher-order gradient
terms. The resulting diode response is controlled by the texture-rigidity
ratio $\Delta_Z/\Delta_\so$ and by the proximity of the triplet channel to
its own condensation threshold, and it differs qualitatively, in both
magnitude and phase-diagram structure, from the mechanisms based on the
band splitting alone \cite{Hasan2024,Bankier2025}.

The paper is organized as follows. Section~\ref{sec:GLSDE} reviews the GL
theory of the diode coefficient for a generic single-component condensate
and establishes the order counting in $\qv$ required for a consistent
result. Section~\ref{sec:pheno} develops the phenomenology: the
symmetry-constrained two-component GL functional of the coupled singlet
and triplet order parameters, its exact reduction to an effective
single-component theory, and the resulting diode efficiency.
Section~\ref{sec:micro} derives all coefficients of the functional from
the microscopic model of an Ising superconductor with a weak Rashba
component, using an extension of the Ref. \cite{Mockli2019} formulation to a
finite Cooper-pair momentum. Section~\ref{sec:results} presents the phase
diagram and the numerical evaluation of the diode efficiency across it.
Section~\ref{sec:discussion} summarizes the results and compares the
triplet-mediated Lifshitz invariants with the conventional ones of the
Rashba and Ising-Rashba models. Technical material is collected in the
appendixes.

%=====================================================================
%=====================================================================
%=====================================================================
%=====================================================================
%=====================================================================
\section{Diode coefficient in Ginzburg-Landau theory}
\label{sec:GLSDE}

We begin with the GL description of a single-component condensate with the
plane-wave ansatz $\Delta(\bm r)=\Delta e^{i\qv\cdot\bm r}$, following
Refs.~\cite{Daido2022,He2022,Yuan2022,Hasan2024}. The free-energy density
is
\begin{equation}
f(\Delta,\qv,\Bv)=\alpha(\qv,\Bv)\,\Delta^2+\beta(\qv,\Bv)\,\Delta^4 ,
\label{eq:GLgeneric}
\end{equation}
where for a fixed direction of $\qv$ the coefficient functions are
expanded as
\begin{equation}
\alpha=\sum_{n\ge0}\alpha_n q^n,
\qquad
\beta=\sum_{n\ge0}\beta_n q^n .
\label{eq:expansion}
\end{equation}
The even-$n$ terms are the conventional condensation and stiffness terms.
The odd-$n$ terms require broken $\mathcal P$ and $\mathcal T$; the
leading one, $\alpha_1q$, is the Lifshitz invariant, and the cubic terms
$\alpha_3q^3$ and $\beta_1q$ are its gradient extensions. Optimizing
Eq.~\eqref{eq:GLgeneric} with respect to $\Delta$ gives the condensation
energy and the supercurrent
\begin{equation}
F(q)=-\frac{\alpha^2(q)}{4\beta(q)}\equiv-\frac{g^2(q)}{4},
\qquad
J(q)=2\,\partial_qF(q),
\label{eq:FandJ}
\end{equation}
where $g(q)=\alpha(q)/\sqrt{\beta(q)}=\sum_ng_nq^n$ is the reduced GL
function. The condensate momentum of the ground state, $q_0$, is fixed by
$J(q_0)=0$; expanding about it,
\begin{equation}
g(q_0+\delta q)=a_0+a_2\,\delta q^2+a_3\,\delta q^3+\cdots ,
\label{eq:aexpansion}
\end{equation}
the critical currents in the two directions and the diode coefficient are
\cite{Hasan2024}
\begin{subequations}\label{eq:etadef}
\begin{align}
&J_{c\pm}=\frac{4a_0^2a_3}{9a_2}\pm\frac{4(-a_0)^{3/2}\sqrt{a_2}}{3\sqrt3},
\\
&\eSDE=\frac{J_{c+}-|J_{c-}|}{J_{c+}+|J_{c-}|}
=\frac{\sqrt{-a_0}\,a_3}{\sqrt3\,a_2^{3/2}} .
\end{align}
\end{subequations}
If only $a_0$ and $a_2$ are kept, $F$ is even in $\delta q$ and the two
critical currents coincide: the Lifshitz invariant alone shifts $q_0$ but
produces no diodicity. To capture the SDE one must work to cubic order in
$\delta q$, which in terms of the original coefficients requires $\alpha$
to order $q^4$ and $\beta$ to order $q^2$, each including the anomalous
odd terms to linear order in $\Bv$. Carrying out the expansion to this
order \cite{Hasan2024},
\begin{equation}
\eSDE=
\frac{2\alpha_2\alpha_3\beta_0-4\alpha_1\alpha_4\beta_0
-\alpha_2^2\beta_1+\alpha_1\alpha_2\beta_2}
{2\sqrt3\,\alpha_2^{5/2}\beta_0}\,\sqrt{-\alpha_0}\, .
\label{eq:etamaster}
\end{equation}
Equation~\eqref{eq:etamaster} makes the order counting explicit and
exposes a structural constraint. For free-fermion values of the normal
coefficients one has the identity $\alpha_2\beta_2=4\alpha_4\beta_0$, both
sides being products of the same two frequency sums (see
Sec.~\ref{sec:quarticmicro}), so the terms proportional to $\alpha_1$
cancel in Eq.~\eqref{eq:etamaster}. The diodicity is therefore controlled
by $\alpha_3$ and $\beta_1$, and any proposed mechanism must be traced
through these coefficients rather than through the Lifshitz invariant
alone. The corrections to this cancellation are of relative order
$1-T/\Tcs$ and are kept in the numerical treatment below.

The utility of this section rests on the fact, established next, that the
coupled singlet-triplet theory of an Ising superconductor in an in-plane
field reduces exactly, at the level of the quadratic terms and
perturbatively in the quartic sector, to the single-component form
\eqref{eq:GLgeneric} with effective coefficient functions
$\Aeff(q,\Bv)$ and $\Beff(q,\Bv)$; Eq.~\eqref{eq:etamaster} then applies
with $\alpha_n\to\Aeff_n$, $\beta_n\to\Beff_n$.

%=====================================================================
%=====================================================================
%=====================================================================
%=====================================================================
%=====================================================================
\section{Phenomenology: coupled singlet and triplet order parameters}
\label{sec:pheno}

In this section we construct the GL functional of the coupled singlet and
field-induced triplet order parameters from symmetry alone, eliminate the
triplet sector, and obtain the diode efficiency in terms of the
phenomenological coefficients. The microscopic values of all coefficients
are derived in Sec.~\ref{sec:micro}; the phenomenology, however, already
contains the key physics and takes us to the structure of the results by
the shortest route. It moreover applies equally to scenarios of unconventional non-BCS superconductivity mediated or assisted by the spin fluctuations, see Refs.~\cite{Das2023,Roy2024,Roy2026}, or by Coulomb interaction \cite{Horhold2023,Sohier2025}.

\subsection{Order parameters and transformation properties}
\label{sec:irreps}

The pairing field is organized in spin space as
$\hat\Delta(\kv)=[\psi(\kv)\sigma_0+\dv(\kv)\cdot\bm\sigma]i\sigma_y$,
with the singlet $\psi$ even and the triplet $\dv$-vector odd in $\kv$.
The dominant channel is the trivial singlet, $\psi(\kv)=\psi$. The
relevant triplet channel is dictated by the conversion physics: the
in-plane field deforms the spin texture of the bands and induces pairs
with the $\dv$-vector
\begin{equation}
\dv(\kv)\propto\Bv\times\gv(\kv),
\label{eq:dinduced}
\end{equation}
where $\gv(\kv)$ is the SOC vector of the normal state,
$H_{\rm so}=\gv(\kv)\cdot\bm\sigma$ with $\gv(-\kv)=-\gv(\kv)$. For the
Ising coupling $\gv=\Delta_\so\hat\gamma(\kv)\zhat$, with
$\hat\gamma(\kv)$ the odd basis function of the crystal normalized as
$\langle\hat\gamma^2\rangle_{\rm FS}=1$, Eq.~\eqref{eq:dinduced} gives an
in-plane $\dv$-vector with the $\hat\gamma$ texture,
\begin{equation}
\dv(\kv)=\hat\gamma(\kv)\big(\eta_x\hat{\bm x}+\eta_y\hat{\bm y}\big),
\label{eq:dvector}
\end{equation}
which transforms according to the $E''$ irreducible representation of
$D_{3h}$ \cite{Mockli2019,Haim2020}. The transformation law of the
doublet $\etav=(\eta_x,\eta_y)$ requires care, and we first fix the axis
convention. With the basis function
$\hat\gamma=\mathrm{sgn}[\cos3\varphi_{\kv}]$ used below, the vertical
mirror planes of the crystal are those of the type $x\to-x$ (the $yz$
plane and its images under the threefold rotation): under $x\to-x$ one
has $\varphi_{\kv}\to\pi-\varphi_{\kv}$, so that
$\cos3\varphi_{\kv}\to-\cos3\varphi_{\kv}$, and simultaneously
$\sigma_z\to-\sigma_z$, leaving the Ising coupling invariant; the $xz$
mirror is not an element of the group in this convention. Although $\dv$
is an axial vector, the basis function $\hat\gamma$ is odd under the
vertical mirrors, and the doublet defined by Eq.~\eqref{eq:dvector}
absorbs this sign: under $x\to-x$ the axial pattern of the unit vectors
$(\hat{\bm x},\hat{\bm y})$ combines with $\hat\gamma\to-\hat\gamma$ to
give $(\eta_x,\eta_y)\to(-\eta_x,\eta_y)$. The doublet therefore
transforms as an in-plane polar vector, identically to $(q_x,q_y)$ and to
the rotated field combination $\zhat\times\Bv=(-B_y,B_x)$, and not as the
axial pair $(B_x,B_y)$; it still spans $E''$, the extra sign reflecting
the mirror-odd scalar factor $\hat\gamma$ in its definition.

The building blocks of the functional and their transformation properties
are then as follows. Under proper rotations about $\zhat$ all in-plane
doublets rotate alike. Under the vertical mirror $x\to-x$,
$\qv\to(-q_x,q_y)$ and $\etav\to(-\eta_x,\eta_y)$ follow the polar
pattern, while the axial field transforms as $\Bv\to(B_x,-B_y)$, so that
$\zhat\times\Bv$ again follows the polar pattern. Under the basal mirror
$\sigma_h$ the in-plane momentum is invariant, while in-plane axial
vectors are odd: $\Bv_\parallel$, $\zhat\times\Bv$, and $\etav$ change
sign ($\hat\gamma$ is $\sigma_h$-even), and any polynomial built from
$\qv$ alone is even. Under time reversal, $\qv\to-\qv$, $\Bv\to-\Bv$,
$\psi\to\psi^*$, $\etav\to\etav^*$.

\subsection{Symmetry-constrained functional}
\label{sec:functional}

Collecting all invariants bilinear in the order parameters to the order
required by Sec.~\ref{sec:GLSDE} (quartic in $q$ on the diagonal, cubic in
$q$ and linear in $B$ in the couplings), the quadratic sector for the
point groups $C_{3v}$ and $C_{\infty v}$ reads
\begin{widetext}
\begin{align}
f_2[\psi,\etav]=&\;\big(\epsilon_0+a_BB^2+\xi^2q^2+\xi_4q^4\big)
|\psi_{\qv}|^2
+\Big[\big(\epsilon_0'+a_B'B^2\big)\delta_{ij}+a_B''B_iB_j
+\xi'^2q^2\delta_{ij}+\xi''^2q_iq_j\Big]\eta^*_{\qv i}\eta_{\qv j}
\nonumber\\
&+2\,\Imm\Big\{\psi^*_{\qv}\Big[
(c_B+c_{B2}q^2)(\zhat\times\Bv)+(c_q+c_{q2}q^2)\,\qv\Big]
\cdot\etav_{\qv}\Big\}
+2\,c_w(\qv)\Ree\big\{\psi^*_{\qv}\,\Bv\cdot\etav_{\qv}\big\}
\nonumber\\
&+\alpha_1^{\rm c}\,\zhat\cdot(\qv\times\Bv)\,|\psi_{\qv}|^2
+\alpha_3^{\rm c}\,q^2\,\zhat\cdot(\qv\times\Bv)\,|\psi_{\qv}|^2 ,
\label{eq:f2}
\end{align}
\end{widetext}
where $c_w(\qv)\propto q^3\cos3\theta_q$ is the warping coupling allowed
in $C_{3v}$, and $\alpha_{1,3}^{\rm c}$ are the anomalous terms of the
pure singlet sector, generated by the band splitting independently of the
triplet channel \cite{Edelstein1996,Hasan2024,Bankier2025}. Every
invariant in Eq.~\eqref{eq:f2} can be checked against the transformation
rules of Sec.~\ref{sec:irreps}; for instance, the coupling
$\Imm\{\psi^*\,\qv\cdot\etav\}$ is invariant because
$\qv$ and $\etav$ share the same rotation and mirror patterns, and it is
$\mathcal T$-even because both the imaginary part and $\qv$ change sign;
the warping term pairs the mirror-odd contraction $\Bv\cdot\etav$ with
the equally mirror-odd threefold harmonic contained in $c_w(\qv)$. 
The third term of Eq.~\eqref{eq:f2} controls the conversion of the singlet pairs into triplet ones, induced by the field and pair momentum, and parametrized by the conversion couplings $c_{B(2)}$ and $c_{q(2)}$, respectively.

The quartic sector, to the order entering
Eq.~\eqref{eq:etamaster}, is
\begin{align}
f_4[\psi,\etav]=&\;\big(\beta_0+\beta_2^{(4)}q^2\big)|\psi|^4
+\beta_1^{\rm c}\,\zhat\cdot(\hat\qv\times\Bv)\,q\,|\psi|^4
\nonumber\\
&+2\,\Imm\Big\{|\psi|^2\psi^*\big[b_B(\zhat\times\Bv)+b_q\,\qv\big]
\cdot\etav\Big\}
\nonumber\\
&+g_1|\psi|^2|\etav|^2+g_2\,\Ree\big\{\psi^{*2}(\etav\cdot\etav)\big\} ,
\label{eq:f4}
\end{align}
where $b_B$ and $b_q$ are the quartic analogs of the conversion couplings
and the $g_{1,2}$ terms contribute only even-in-$q$ renormalizations at
the order considered.

For the mirror-symmetric group $D_{3h}$ a strong selection rule applies.
Any bilinear coupling $\psi^*T_i(\qv)\eta_i$ with $T_i$ built solely from
the components of the in-plane momentum is $\sigma_h$-even times the
$\sigma_h$-odd doublet $\etav$, and therefore vanishes to all orders in
$\qv$; the same would hold for an extra 
terms containing the $\sigma_h$-even  component $B_z$.
Consequently, in $D_{3h}$,
\begin{equation}
c_q=c_{q2}=c_w=b_q=0,\qquad \alpha_{1,3}^{\rm c}=\beta_1^{\rm c}=0 ,
\label{eq:D3hselection}
\end{equation}
and only the field couplings $c_B$, $b_B$ survive: the in-plane field
induces the $E''$ triplet, but no Lifshitz invariant and no diode effect
can be generated at any order in the gradients, in accord with the
$\sigma_h\mathcal T$ argument of the Introduction. An out-of-plane field
induces no triplets at all, since $\Bv\times\gv=0$ for
$\Bv\parallel\gv\parallel\zhat$. The entire triplet route to the SDE
therefore switches on only when the basal mirror is broken, and all the
mirror-odd couplings in Eqs.~\eqref{eq:f2} and \eqref{eq:f4} are
proportional to the Rashba component of the SOC. Microscopically
(Sec.~\ref{sec:micro}) they obey the kinematic locking
\begin{equation}
\frac{c_q}{c_B}=\frac{c_{q2}}{c_{B2}}=\frac{b_q}{b_B}=-\frac{\alpha_R}{2},
\label{eq:locking}
\end{equation}
with $\alpha_R$ the Rashba velocity, a relation that is independent of the
interaction constants, the temperature, and the Fermi energy.

\subsection{Elimination of the triplet sector}
\label{sec:elimination}

The triplet doublet is massive in the parameter range of interest
($\epsilon_0'>0$) and can be integrated out exactly at the quadratic
level. Writing the couplings of Eq.~\eqref{eq:f2} as
$2\Imm\{\psi^*\bm u\cdot\etav\}$ with the real vector
\begin{equation}
\bm u(\qv,\Bv)=(c_B+c_{B2}q^2)(\zhat\times\Bv)+(c_q+c_{q2}q^2)\,\qv ,
\label{eq:uvector}
\end{equation}
and writing the triplet mass term as a 
quadratic form, $\eta_i^*M_{ij}\eta_j$, the 
optimization
with respect to $\etav^*$ gives
\begin{equation}
\etav=-\frac{i}{M}\,\bm u\,\psi
\label{eq:etasol}
\end{equation}
for isotropic $M$: the induced triplet is phase shifted by $\pi/2$
relative to the singlet, which is the GL image of the $s+if$ structure of
the field-driven state \cite{Mockli2019}. Substituting back,
\begin{equation}
\Aeff(\qv,\Bv)=\alpha_{\rm s}(\qv,\Bv)
-\bm u^{\top}(\qv,\Bv)\,M^{-1}(\qv)\,\bm u(\qv,\Bv) ,
\label{eq:Aeffdef}
\end{equation}
where $\alpha_{\rm s}$ collects the singlet-diagonal terms of
Eq.~\eqref{eq:f2}. With the anisotropic triplet form,
$M=M_\parallel$ on $\hat\qv$ and $M_\perp$ on $\zhat\times\hat\qv$, with
$M_\perp=\epsilon_0'+\xi'^2q^2$ and
$M_\parallel=M_\perp+\xi''^2q^2$, the two projections of $\bm u$ are
[using $(\zhat\times\Bv)\cdot\hat\qv=-\zhat\cdot(\hat\qv\times\Bv)$ and
$(\zhat\times\Bv)\cdot(\zhat\times\hat\qv)=\Bv\cdot\hat\qv$]
\begin{equation}
u_\parallel=(c_q+c_{q2}q^2)\,q-c_B\,\zhat\cdot(\hat\qv\times\Bv),
\qquad
u_\perp=c_B\,\Bv\cdot\hat\qv ,
\label{eq:uproj}
\end{equation}
so that
\begin{equation}
\Aeff=\alpha_{\rm s}
-\frac{u_\parallel^2}{M_\parallel}-\frac{u_\perp^2}{M_\perp}.
\label{eq:Aeffproj}
\end{equation}
The Lifshitz cross term thus resides in the longitudinal projection,
which carries the stiffness combination $\xi'^2+\xi''^2$. Expanding
Eq.~\eqref{eq:Aeffproj} in $q$ with
$M_\parallel^{-1}=\epsilon_0'^{-1}
[1-(\xi'^2+\xi''^2)q^2/\epsilon_0'+\cdots]$,
$M_\perp^{-1}=\epsilon_0'^{-1}
(1-\xi'^2q^2/\epsilon_0'+\cdots)$, and denoting
$B_\perp=\zhat\cdot(\hat\qv\times\Bv)$, the coefficients of the effective
single-component theory follow term by term:
\begin{subequations}
\label{eq:Aeffcoeffs}
\begin{align}
\Aeff_0&=\epsilon_0+a_BB^2-\frac{c_B^2B^2}{\epsilon_0'} ,
\\
\Aeff_1&=\alpha_1^{\rm c}+\frac{2c_qc_B}{\epsilon_0'}\,B_\perp ,
\label{eq:A1}
\\
\Aeff_2&=\xi^2-\frac{c_q^2}{\epsilon_0'}
+\frac{c_B^2B^2\xi'^2}{\epsilon_0'^2}
+\frac{c_B^2\xi''^2}{\epsilon_0'^2}\,B_\perp^2 ,
\\
\Aeff_3&=\alpha_3^{\rm c}
+\frac{2(c_{q2}c_B+c_{B2}c_q)}{\epsilon_0'}\,B_\perp
-\frac{2c_qc_B\,(\xi'^2+\xi''^2)}{\epsilon_0'^2}\,B_\perp ,
\label{eq:A3}
\\
\Aeff_4&=\xi_4+\frac{(\xi'^2+\xi''^2)\,c_q^2}{\epsilon_0'^2}
-\frac{2c_{q2}c_q}{\epsilon_0'} .
\end{align}
\end{subequations}
The odd coefficients \eqref{eq:A1} and \eqref{eq:A3} are the Lifshitz
invariant and its cubic extension generated by the field-induced triplet
channel: they are bilinear in the two conversion couplings, odd in $B$,
and carry the triplet-channel pole $1/\epsilon_0'$. In the quartic
sector, inserting Eq.~\eqref{eq:etasol} into Eq.~\eqref{eq:f4},
\begin{subequations}
\label{eq:Beffcoeffs}
\begin{align}
\Beff_0&=\beta_0-\frac{2b_Bc_BB^2}{\epsilon_0'},
\qquad
\Beff_2=\beta_2^{(4)}-\frac{2b_qc_q}{\epsilon_0'},
\\
\Beff_1&=\beta_1^{\rm c}
+\frac{2\big(b_Bc_q-b_qc_B\big)}{\epsilon_0'}\,B_\perp .
\label{eq:B1}
\end{align}
\end{subequations}
Equations \eqref{eq:Aeffcoeffs} and \eqref{eq:Beffcoeffs} accomplish the
reduction announced in Sec.~\ref{sec:GLSDE}: the coupled theory is now a
single-component GL theory with coefficient functions $\Aeff$, $\Beff$,
and the diode coefficient follows from Eq.~\eqref{eq:etamaster}. Note the
consistency checks: all odd-in-$q$ coefficients are odd in $B$ and vanish
when either coupling family vanishes; in $D_{3h}$, where
Eq.~\eqref{eq:D3hselection} holds, every odd term disappears; and all
triplet-mediated terms are cut off by the triplet mass.

\subsection{Covariant structure and the origin of diodicity}
\label{sec:covariant}

The locking \eqref{eq:locking} organizes the anomalous terms into a
compact form. To linear order in $\alpha_R$ the couplings depend on $\qv$
and $\Bv$ only through the combination
\begin{equation}
\bv=\Bv-\frac{\alpha_R}{2}\,(\qv\times\zhat) ,
\label{eq:bdef}
\end{equation}
whose microscopic origin is derived in Sec.~\ref{sec:lemma}: a condensate
moving with momentum $\qv$ experiences, through the Rashba coupling, an
effective in-plane field that adds to the applied one. In particular,
$\bm u=c_B(q^2)\,\zhat\times\bv$, so that
$\bm u\cdot\bm u=c_B^2(q^2)\,b^2$ with
\begin{equation}
b^2=B^2+\alpha_R\,\zhat\cdot(\qv\times\Bv)+\frac{\alpha_R^2q^2}{4} ,
\label{eq:b2}
\end{equation}
so the cross term of $b^2$ has precisely the Lifshitz-invariant structure.
This has an important structural consequence: if the free energy were a
function of $\bv$ alone, it would be symmetric under the inversion of
$\delta q$ about the shifted minimum, and the diode coefficient would
vanish, in close analogy with the approximate inversion symmetry of the
Rashba superconductor \cite{Hasan2024}. Diodicity requires terms that
break the $\bv$ covariance. Two such terms are present in
Eqs.~\eqref{eq:Aeffcoeffs} and \eqref{eq:Beffcoeffs}: the explicit
$q^2$ dependence of the couplings and masses generated by the scalar
Doppler shift of the quasiparticles, and the triplet stiffness term
$\propto\xi'^2+\xi''^2$ in Eq.~\eqref{eq:A3}. The microscopic calculation below
confirms that the entire diode response of the model arises from the
interplay of the scalar Doppler effect with the covariant field $\bv$.

\subsection{Diode efficiency}
\label{sec:phenoeta}

Substituting Eqs.~\eqref{eq:Aeffcoeffs} and \eqref{eq:Beffcoeffs} into
Eq.~\eqref{eq:etamaster} and using the free-coefficient identity
$\alpha_2\beta_2=4\alpha_4\beta_0$ to drop the $\Aeff_1$ terms at leading
order, the triplet-mediated part of the diode efficiency reads
\begin{align}
\delta\eSDE^{\rm trip}
=\frac{\sqrt{-\Aeff_0}}{\sqrt3\,\Aeff_2^{3/2}}
\Big[&\frac{2(c_{q2}c_B+c_{B2}c_q)}{\epsilon_0'}
-\frac{2c_qc_B\,(\xi'^2+\xi''^2)}{\epsilon_0'^2}
\nonumber\\[-2pt]
&-\frac{\Aeff_2}{\Beff_0}\,
\frac{(b_Bc_q-b_qc_B)}{\epsilon_0'}\Big]B_\perp ,
\label{eq:etatrip}
\end{align}
to be evaluated with the microscopic coefficients of
Sec.~\ref{sec:micro}. Every term is linear in $B$ and linear in
$\alpha_R$, as required by symmetry; the relative weight of the three
contributions, and the sign of the net effect relative to the
singlet-sector terms $\alpha_3^{\rm c}$, $\beta_1^{\rm c}$, is a
quantitative question that cannot be settled phenomenologically.
It is answered by the microscopic theory, to which we now turn.

%=====================================================================
%=====================================================================
%=====================================================================
%=====================================================================
%=====================================================================
\section{Microscopic theory}
\label{sec:micro}

\subsection{Model and quasiclassical formulation}
\label{sec:model}

The normal state is described by
\begin{equation}
H_0=\sum_{\kv}\xi(\kv)c^\dagger_{\kv\sigma}c_{\kv\sigma}
+\sum_{\kv,\sigma\sigma'}\big[\gv(\kv)-\Bv\big]
\cdot\bm\sigma_{\sigma\sigma'}c^\dagger_{\kv\sigma}c_{\kv\sigma'} ,
\label{eq:H0}
\end{equation}
with the SOC vector of the Ising-Rashba form
\begin{equation}
\gv(\kv)=\Delta_\so\hat\gamma(\kv)\,\zhat+\alpha_R(\kv\times\zhat),
\qquad \alpha_Rk_F\ll\Delta_\so\ll E_F ,
\label{eq:gvector}
\end{equation}
and we absorb $g\mu_B/2$ into $\Bv$ and set $\hbar=k_B=1$. For
concreteness we use the $K$-model basis function
$\hat\gamma=\mathrm{sgn}[\cos3\varphi_{\kv}]$, appropriate for Fermi pockets
away from the zone center \cite{Mockli2020,Haim2020}; the modifications
for the $\Gamma$-model are summarized in Appendix~\ref{app:warping}. The
pairing interaction contains the trivial singlet channel with coupling
constant defining the transition temperature $\Tcs$ and the $E''$ triplet
channel of Eq.~\eqref{eq:dvector} with the bare transition temperature
$\Tct<\Tcs$ \cite{Mockli2019,Haim2020}. The natural dimensionless
parameters are
\begin{equation}
\delta=\frac{\alpha_R}{v_F},\qquad
y=\frac{\Delta_\so}{\pi T},\qquad
t=1-\frac{T}{\Tcs},\qquad
\tau=\frac{\Tct}{\Tcs} .
\end{equation}

In the quasiclassical regime the linearized gap equations follow from the
Eilenberger equations \cite{Mockli2020}. At zero Cooper-pair momentum
they read $\omega_nf_0=\psi+i\bm f\cdot\Bv$ and
$\omega_n\bm f=\dv+if_0\Bv+\gv\times\bm f$, where $f_0$ and $\bm f$ are
the singlet and triplet pairing correlations at Matsubara frequency
$\omega_n$. Two structural facts orient the whole calculation. First, at
$\Bv=0$ the equations decouple $f_0$ from $\bm f$ for any $\gv(\kv)$ and
any $\qv$: within the leading quasiclassical order there is no gradient
conversion between singlets and triplets, so all gradient couplings in
Eq.~\eqref{eq:f2} must be proportional to the band-resolved asymmetries
generated by the Rashba velocity, in accord with
Eq.~\eqref{eq:locking}. Second, the term $if_0\Bv$ converts singlet into
triplet correlations with no Fermi-energy suppression, the amplitude
being limited only by the texture rigidity: this is the asymmetry between
$c_q\propto\alpha_R$ and the Fermi-energy-independent $c_B$ announced in
the Introduction.

\subsection{Finite momentum: the covariant Doppler substitution}
\label{sec:lemma}

The free energy of the model is generated by the standard fermionic trace
formula [Appendix~\ref{app:kernels}, Eq.~\eqref{eq:traceformula}], in
which the pairing field $\Delta e^{i\qv\cdot\bm r}$ attaches the momenta
$\kv+\qv/2$ and $-\kv+\qv/2$ to the two fermion lines of every loop.
Expanding the single-particle Hamiltonian at the shifted momenta gives,
for the scalar part, the Doppler shift
$\xi_{\pm\kv+\qv/2}=\xi_{\kv}\pm\frac12\bm v_F\cdot\qv$, and for the spin
part
\begin{equation}
\gv(\pm\kv+\qv/2)=\pm\gv(\kv)+\frac{\alpha_R}{2}(\qv\times\zhat)
\pm\frac{\Delta_\so}{2}(\qv\cdot\nabla_{\kv}\hat\gamma)\,\zhat ,
\label{eq:shifted}
\end{equation}
since the Rashba term is linear in $\kv$ while the Ising term is odd. For
the $K$-model the last term vanishes almost everywhere on the Fermi
surface; for the $\Gamma$-model it produces the warping coupling
(Appendix~\ref{app:warping}). The momentum-even part of the Rashba shift
is common to both lines and combines with the Zeeman term into the
covariant field of Eq.~\eqref{eq:bdef}: the particle line carries the
spin structure $(\gv_{\kv}-\bv)\cdot\bm\sigma$ and the partner line
$-(\gv_{\kv}+\bv)\cdot\bm\sigma$, which is exactly the spin structure of
the Gor'kov blocks of the zero-momentum problem with $\Bv\to\bv$
\cite{Mockli2019}. Hence, to all orders of the expansion of the trace
formula, the finite-momentum free energy of the $K$-model follows from
the zero-momentum theory of Refs, \cite{Mockli2019,Mockli2020} by the two replacements
\begin{subequations}\label{eq:replacements}
\begin{align}
\Bv\to\bv=\Bv-\frac{\alpha_R}{2}(\qv\times\zhat),
\\
\omega_n\to s_n=\omega_n+\tfrac i2\,\mathrm{sgn}(\omega_n)\,
\bm v_F\cdot\qv
\end{align}
\end{subequations}
followed by the Fermi-surface angular average. Equation
\eqref{eq:replacements} is the central technical result of this section. It
proves the locking \eqref{eq:locking} for every vertex of the theory,
quadratic and quartic, since $\qv$ and $\Bv$ enter each vertex only
through $\bv$; corrections are of relative order
$\Delta_R^2/\Delta_\so^2$, with $\Delta_R=\alpha_Rk_F$, or of warping
origin. For the diode geometry, $\qv=q\hat{\bm x}$ and
$\Bv=B\hat{\bm y}$, the covariant field is collinear with $\Bv$ and
reduces to the scalar $b(q)=B+\alpha_Rq/2$, with $b^2$ given by
Eq.~\eqref{eq:b2}. As shown below, the conversion couples the singlet to
the doublet projection along $\zhat\times\hat\bv$, i.e., it contracts
$\etav$ with $\qv$ and with $\zhat\times\Bv$, confirming the polar
transformation law assigned to $\etav$ in Sec.~\ref{sec:irreps}.

\subsection{Matsubara sums: a single generating function}
\label{sec:technology}

All frequency sums of the theory reduce to the pair-breaking function
introduced in Ref.~\cite{Mockli2019},
\begin{equation}
C(y)=\Ree\,\psi\Big(\frac12+\frac{iy}{2}\Big)-\psi\Big(\frac12\Big) ,
\label{eq:Cdef}
\end{equation}
its derivatives
\begin{subequations}\label{eq:Cderivs}
\begin{align}
C'(y)=-\frac12\Imm\,\psi^{(1)}\Big(\frac12+\frac{iy}2\Big),
\\
C''(y)=-\frac14\Ree\,\psi^{(2)}\Big(\frac12+\frac{iy}2\Big),
\end{align}
\end{subequations}
and the constants $\zeta(3)$, $\zeta(5)$. The basic identity, derived by
partial fractions in Appendix~\ref{app:sums}, is
\begin{equation}
\pi T\sum_{n}\frac{\Delta^2}{|\omega_n|(\omega_n^2+\Delta^2)}=C(y),
\qquad y=\frac{\Delta}{\pi T} ,
\label{eq:Cidentity}
\end{equation}
with the asymptote $C(y\gg1)\simeq\ln(2e^{\gamma_E}y)$. Differentiation
with respect to $\Delta^2$ generates the derivative identities collected
in Appendix~\ref{app:sums}; the one used most often below is
\begin{equation}
\pi T\sum_n\frac{2|\omega_n|(\omega_n^2-3\Delta^2)}
{(\omega_n^2+\Delta^2)^3}=\frac{C''(y)}{(\pi T)^2} .
\label{eq:Cppidentity}
\end{equation}
The Doppler averages follow from the expansion
\begin{equation}
\big\langle\Ree\,K(\omega_n+iu)\big\rangle_\theta
=K-\frac{\langle u^2\rangle}{2}K''
+\frac{\langle u^4\rangle}{24}K''''+\cdots,
\label{eq:dopplerexp}
\end{equation}
with $u=\frac12v_Fq\cos\Theta$, $\langle u^2\rangle=v_F^2q^2/8$, and
$\langle u^4\rangle=3v_F^4q^4/128$; odd powers of $u$ average to zero
except against the warping harmonics of $\hat\gamma$
(Appendix~\ref{app:warping}).

\subsection{Quadratic sector}
\label{sec:quadmicro}

Solving the linearized equations at finite $\qv$ and in-plane $\bv$
(Appendix~\ref{app:kernels}) and feeding the solutions into the
self-consistency conditions of the two channels yields the quadratic form
$F_2/2N_0=\alpha_s\psi^2+\alpha_t\eta_c^2+2\alpha_{st}\psi\eta_c$, where
$N_0$ is the density of states per spin, $\eta_c$ is the triplet
component along $\zhat\times\hat{\bv}$ (the induced combination), and
we use the real-field convention of Ref.~\cite{Mockli2019} in which the
factor $i$ of the $s+if$ state is made explicit (for the field along
$\hat{\bm x}$, as in that reference, the induced component is
$d_y=i\eta_c\hat\gamma$).
The entries are Matsubara sums over four per-frequency kernels,
\begin{gather}
K_s=\frac{s^2+\Delta_\so^2}{sW},\qquad
K_{t\perp}=\frac{s}{s^2+\Delta_\so^2}
+\frac{\Delta_\so^2b^2}{sW(s^2+\Delta_\so^2)},
\nonumber\\
K_{t\parallel}=\frac{s}{W},\qquad
K_{st}=\frac{\Delta_\so b}{sW},
\qquad W=s^2+\Delta_\so^2+b^2 ,
\label{eq:kernels}
\end{gather}
according to
\begin{align}
\alpha_s&=\ln\frac{T}{\Tcs}
+\pi T\sum_n\Big\langle\frac{1}{|\omega_n|}-\Ree K_s\Big\rangle_\theta,
\nonumber\\
\alpha_t&=\ln\frac{T}{\Tct}
+\pi T\sum_n\Big\langle\frac{1}{|\omega_n|}-\Ree K_{t\perp}
\Big\rangle_\theta,
\nonumber\\
\alpha_{st}&=-\pi T\sum_n\big\langle\Ree K_{st}\big\rangle_\theta .
\label{eq:entries}
\end{align}
The component of $\etav$ along $\hat{\bv}$ does not couple to the singlet
at the quadratic level and is governed by $K_{t\parallel}$.

At zero Doppler shift ($u=0$) the sums are elementary
(Appendix~\ref{app:sums}) and give, with
$\tilde\rho^2=\Delta_\so^2+b^2$ and $\rho=\tilde\rho/\pi T$,
\begin{align}
\alpha_s&=\ln\frac{T}{\Tcs}+\frac{b^2}{\tilde\rho^2}C(\rho),
\qquad
\alpha_{st}=-\frac{b\Delta_\so}{\tilde\rho^2}C(\rho),
\nonumber\\
\alpha_t&=\ln\frac{T}{\Tct}+\frac{\Delta_\so^2}{\tilde\rho^2}C(\rho) ,
\label{eq:MKmatrix}
\end{align}
which reproduces the zero-momentum matrix of Ref.~\cite{Mockli2019} with
$B\to b$, a nontrivial check of the formalism. Equation
\eqref{eq:MKmatrix} contains, at finite momentum, the exact
field-dependent triplet mass
\begin{equation}
\epsilon_0'(T,B)=\ln\frac{T}{\Tct}
+\frac{\Delta_\so^2}{\Delta_\so^2+B^2}\,
C\Big(\frac{\sqrt{\Delta_\so^2+B^2}}{\pi T}\Big) ,
\label{eq:massfull}
\end{equation}
which decreases with field: the spin-orbit limiting of the $E''$ doublet
is weighted by $\Delta_\so^2/(\Delta_\so^2+B^2)$ and is quenched at
$B\gtrsim\Delta_\so$. This softening is the finite-momentum image of the
divergence of the critical field at $T\to\Tct$ found in
Ref.~\cite{Mockli2019} and plays the central role in the results below.

The Doppler expansion \eqref{eq:dopplerexp} applied to the kernels
\eqref{eq:kernels}, with the identities of Appendix~\ref{app:sums},
yields the gradient coefficients. The singlet sector reproduces the
free-fermion stiffnesses
\begin{equation}
\xi^2=\frac{7\zeta(3)v_F^2}{32\pi^2T^2},
\qquad
\xi_4=-\frac{93\zeta(5)v_F^4}{2048\pi^4T^4} ,
\label{eq:xifree}
\end{equation}
in agreement with Ref.~\cite{Hasan2024}, plus the field term
$b^2C(\rho)/\tilde\rho^2$ carrying its own Doppler correction. The
conversion vertex acquires the momentum dependence
\begin{equation}
\alpha_{st}(q)=-\frac{b(q)\,\Delta_\so}{\tilde\rho^2}
\Big[C(\rho)-\frac{v_F^2q^2}{16\pi^2T^2}
\Big(\frac{7\zeta(3)}{2}-C''(\rho)\Big)\Big] ,
\label{eq:Astq}
\end{equation}
so that both cubic vertices of Eq.~\eqref{eq:f2} descend from a single
kernel derivative,
$c_{B2}/c_B=c_{q2}/c_q
=-\tfrac{v_F^2}{16\pi^2T^2}[\tfrac72\zeta(3)-C'']/C$,
as anticipated by the covariant structure. The triplet stiffness comes
out in closed form,
\begin{equation}
\xi'^2=\frac{v_F^2\,C''(y)}{16\pi^2T^2}
\;\xrightarrow[\;T\ll\Delta_\so\;]{}\;
-\frac{v_F^2}{16\,\Delta_\so^2}
\label{eq:xitriplet}
\end{equation}
and is negative, since $C''<0$. The interpretation is transparent: the
spin-orbit coupling acts on the $E''$ pair as a pair-breaking field, so
the triplet susceptibility is maximal at a finite Doppler detuning, in
direct analogy with the Fulde-Ferrell-Larkin-Ovchinnikov mechanism
\cite{FuldeFerrell1964,LarkinOvchinnikov1964}. The channel remains
stable because it is massive, and the stiffness correction to the
observables is small, of relative order $(T/\Delta_\so)^2$, but its sign
matters for the composition of $\Aeff_3$ in Eq.~\eqref{eq:A3}.

\subsection{Quartic sector}
\label{sec:quarticmicro}

The substitution \eqref{eq:replacements} applies verbatim to the quartic
term of the trace formula, so the zero-momentum quartic free energy of
Ref.~\cite{Mockli2019} extends to finite momentum by $B\to b(q)$, with
Doppler corrections entering only the even-in-$q$ sector at the order
required by Eq.~\eqref{eq:etamaster}. In the variables
$P=b\psi-\Delta_\so\eta_c$ and $Q=\Delta_\so\psi+b\eta_c$, which are the
gap projections transverse and parallel to the effective spin field of
the split bands,
\begin{equation}
\frac{F_4}{2N_0}=\beta_1P^2Q^2+\beta_2P^4+\beta_3Q^4 ,
\label{eq:F4}
\end{equation}
and the coefficients, converted from the polygamma forms of
Ref.~\cite{Mockli2019} into the $C$ family using
Eq.~\eqref{eq:Cderivs}, are
\begin{equation}
\beta_1=\frac{\rho C'(\rho)-C(\rho)}{\tilde\rho^6},
\quad
\beta_2=\frac{\rho^2C''(\rho)}{4\tilde\rho^6},
\quad
\beta_3=\frac{7\zeta(3)}{8\pi^2\Tcs^2\tilde\rho^4} ,
\label{eq:betasC}
\end{equation}
evaluated at $T=\Tcs$ as appropriate for the GL regime. The structure
promised in the abstract is now explicit: the entire two-channel
functional, masses, conversion vertices, stiffnesses, and quartic terms,
is generated by the single function $C$ with its first two derivatives,
together with the free-fermion constants. Expanding Eq.~\eqref{eq:F4} at
small $\eta_c$ gives the quartic conversion vertex in closed form,
\begin{align}
&\frac{\partial(F_4/2N_0)}{\partial\eta_c}\bigg|_{\eta_c=0}\nonumber \\
&=2b\Delta_\so\big[\beta_1(b^2-\Delta_\so^2)-2\beta_2b^2
+2\beta_3\Delta_\so^2\big]\psi^3
\nonumber\\
&\xrightarrow[\;b\to0\;]{}\;
2b\,\Delta_\so^3\,(2\beta_3-\beta_1)\,\psi^3 ,
\label{eq:quarticvertex}
\end{align}
identifying $b_B=2\Delta_\so^3(2\beta_3-\beta_1)$ and, through the
locking, $b_q=-\tfrac{\alpha_R}{2}b_B$. The anomalous quartic term of the
pure singlet sector follows from the $b$ dependence at $\eta_c=0$,
\begin{equation}
\frac{F_4}{2N_0}\bigg|_{\eta_c=0}
=\big[\beta_1\Delta_\so^2b^2+\beta_2b^4+\beta_3\Delta_\so^4\big]\psi^4 ,
\label{eq:F4singlet}
\end{equation}
whose derivative with respect to the invariant
$\alpha_R\zhat\cdot(\qv\times\Bv)$ inside $b^2$ produces
$\beta_1^{\rm c}$ within the model; the Doppler broadening of the aligned
kernel reproduces the free-fermion gradient coefficient
$\beta_2^{(4)}$ of Ref.~\cite{Hasan2024}.

\subsection{Effective functional and the anomalous terms}
\label{sec:anatomy}

Eliminating $\eta_c$ exactly at the quadratic level, with
$r(q)=\eta_c/\psi=-\alpha_{st}/\alpha_t$,
\begin{align}
\Aeff(q,B)&=\alpha_s-\frac{\alpha_{st}^2}{\alpha_t},
\nonumber\\
\Beff(q,B)&=\beta_1\bar P^2\bar Q^2+\beta_2\bar P^4+\beta_3\bar Q^4,
\label{eq:elimexact}
\end{align}
with $\bar P=b-\Delta_\so r$ and $\bar Q=\Delta_\so+br$. Every anomalous
coefficient acquires a two-part anatomy: a direct part from the $b(q)$
dependence of the singlet sector, and a collective part from the $b(q)$
dependence of the conversion vertex and the triplet mass. For the
Lifshitz invariant, differentiating Eq.~\eqref{eq:elimexact} with respect
to the invariant $\ell=\alpha_R\zhat\cdot(\qv\times\Bv)$ at weak field
gives the exact result
\begin{equation}
\Aeff_1=\alpha_R\,\zhat\cdot(\hat\qv\times\Bv)\,
\frac{C(y)}{\Delta_\so^2}
\Big[1-\frac{C(y)}{\epsilon_0'}\Big],
\label{eq:A1anatomy}
\end{equation}
with $\epsilon_0'=\ln(T/\Tct)+C(y)$. The two conversion paths interfere
destructively: the collective contribution carries the opposite sign, the
ratio of the two parts is $C/\epsilon_0'\le1$ at weak field, and the
invariant vanishes identically at $T=\Tct$. The physical origin of the
sign is Eq.~\eqref{eq:etasol}: the triplet induced through the collective
mode is shifted in phase by $\pi/2$, and its back action on the singlet
acquires a factor $i^2=-1$ relative to the direct virtual-pair
contribution. In the quartic sector the pattern is reversed: substituting
$r$ into Eq.~\eqref{eq:elimexact} amplifies the misaligned projection
$\bar P$ regardless of the phase, and the quartic anomalous coefficient is
enhanced,
\begin{equation}
\Beff_1\propto\alpha_R\Big[\Delta_\so^2\beta_1
\Big(1+\frac{C\Delta_\so^2}{\tilde\rho^2\epsilon_0'}\Big)^{2}
+\Delta_\so^4\frac{\partial\beta_3}{\partial b^2}+\cdots\Big] .
\label{eq:B1anatomy}
\end{equation}
The opposite interference patterns in the quadratic and quartic anomalous
sectors determine the structure of the results in
Sec.~\ref{sec:results}. At strong fields the anatomy is controlled by
Eq.~\eqref{eq:massfull}: the collapse of the triplet mass promotes the
collective contributions to the dominant ones.

%=====================================================================
%=====================================================================
%=====================================================================
%=====================================================================
%=====================================================================
\section{Results}
\label{sec:results}

The theory defined by Eqs.~\eqref{eq:entries}, \eqref{eq:elimexact}, and
\eqref{eq:F4} contains no free parameters beyond
$(\Delta_\so/\Tcs,\;\Tct/\Tcs,\;\delta)$ and can be evaluated without
further approximation: for each $q$ the free energy is minimized over the
two order-parameter amplitudes, the supercurrent $J(q)=2\partial_qF$ is
computed, and the diode coefficient follows from the extrema of $J$. We
use $\Delta_\so=15\,\Tcs$, representative of the Ising scale of the
transition-metal dichalcogenides, and $\delta=0.1$; the efficiency is
linear in $\delta$ to a fraction of a percent in the plotted range, and
we report $\eSDE/\delta$. The numerical procedure and its convergence
checks are described in Appendix~\ref{app:numerics}; in particular,
$\eSDE$ vanishes at $\alpha_R=0$, as required by the
symmetry analysis, and the perturbative chain of
Secs.~\ref{sec:GLSDE}-\ref{sec:pheno} was verified against the exact
evaluation, agreeing at the few-percent level near $\Tcs$ and degrading
gradually with $t$ and with the triplet admixture, as expected for a
leading-order expansion.

\begin{figure*}[t]
\centering
\includegraphics[width=0.98\textwidth]{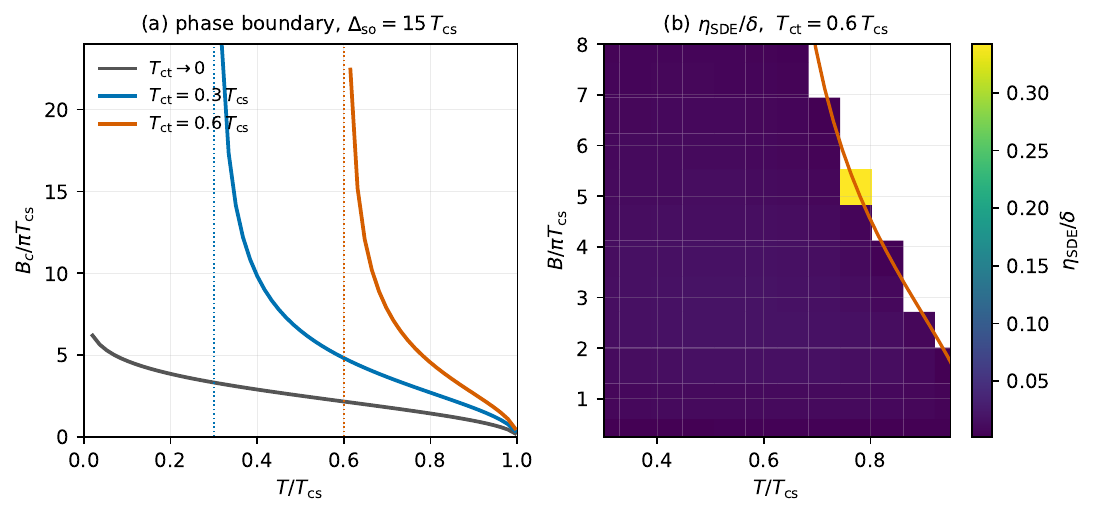}
\caption{(a) Phase boundary $B_c(T)$ of the coupled singlet-triplet
system, obtained from the vanishing of the lowest eigenvalue of the
quadratic form \eqref{eq:MKmatrix}, for three strengths of the triplet
channel. The attractive $E''$ channel enhances $B_c$ and drives its
divergence as $T\to\Tct$ (dotted verticals), consistent with
Ref.~\cite{Mockli2019}. (b) Diode efficiency $\eSDE/\delta$ across the
$(T,B)$ plane for $\Tct=0.6\,\Tcs$, with the phase boundary overlaid.
The efficiency is maximal in a ridge adjacent to the triplet-enhanced
critical field, reaching values an order of magnitude above the low-field
ones. Parameters: $\Delta_\so=15\,\Tcs$, $\delta=0.1$.}
\label{fig:phase}
\end{figure*}

\begin{figure*}[t]
\centering
\includegraphics[width=0.98\textwidth]{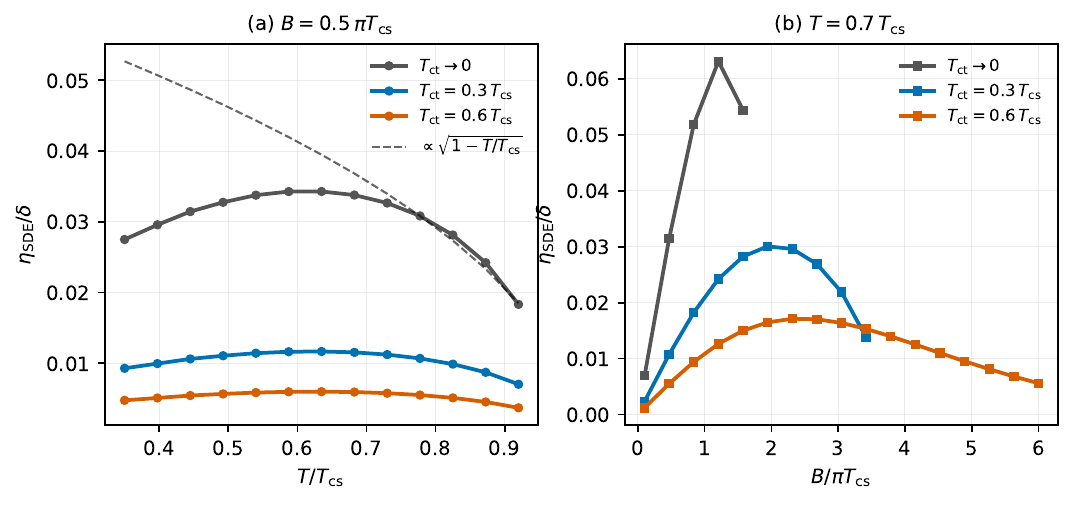}
\caption{(a) Temperature dependence of the diode efficiency at fixed
$B=0.5\,\pi\Tcs$ for three strengths of the triplet channel. The dashed
line is the $\sqrt{1-T/\Tcs}$ envelope of the GL regime, normalized at
$T=0.8\,\Tcs$. (b) Field dependence at $T=0.7\,\Tcs$. The singlet-only
response is cut off at its critical field, while the triplet channel
extends the diode regime several-fold, with the maximum shifted toward
the enhanced $B_c$.}
\label{fig:eta}
\end{figure*}

\begin{figure}[t]
\centering
\includegraphics[width=0.85\columnwidth]{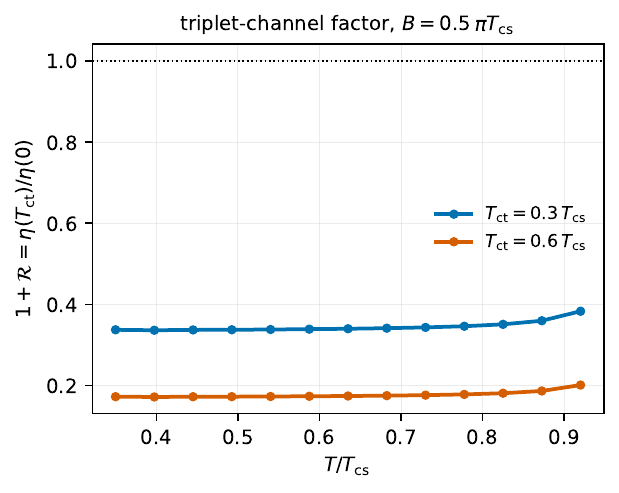}
\caption{Triplet-channel factor
$1+\mathcal R(T)=\eSDE(\Tct)/\eSDE(\Tct\to0)$ at $B=0.5\,\pi\Tcs$. At
weak fields the collective conversion path interferes destructively with
the direct one and suppresses the efficiency by a nearly
temperature-independent factor, in accord with the anatomy of
Eqs.~\eqref{eq:A1anatomy} and \eqref{eq:B1anatomy}.}
\label{fig:R}
\end{figure}

Figure~\ref{fig:phase}(a) shows the phase boundary of the coupled system.
The triplet channel enhances the in-plane critical field and drives its
divergence as $T\to\Tct$: at strong fields the spin-orbit limiting of the
$E''$ doublet is quenched [Eq.~\eqref{eq:massfull}], and
superconductivity survives in a strongly mixed $s+if$ form. The diode
efficiency across the same plane, Fig.~\ref{fig:phase}(b), correlates
directly with this structure: the ridge of maximal $\eSDE$ runs parallel
to and just inside the enhanced $B_c(T)$, in the region where the
condensate exists only by virtue of the triplet admixture. In this regime
the diode effect is a genuinely triplet-driven phenomenon rather than a
correction to the singlet response.

Figure~\ref{fig:eta} quantifies the two field regimes. At weak fields,
panel (a), the efficiency of all three systems follows the
$\sqrt{1-T/\Tcs}$ envelope characteristic of the GL regime, and the
triplet channel rescales the amplitude by a nearly temperature-independent
factor smaller than unity: $1+\mathcal R\simeq0.34$ for
$\Tct=0.3\,\Tcs$ and $\simeq0.17$ for $\Tct=0.6\,\Tcs$ at
$B=0.5\,\pi\Tcs$ (Fig.~\ref{fig:R}). This suppression is the destructive
interference derived in Sec.~\ref{sec:anatomy}: the quadratic-sector
anatomy \eqref{eq:A1anatomy} would give
$\ln(T/\Tct)/[\ln(T/\Tct)+C]\simeq0.21$ for $\Tct=0.3\,\Tcs$ at
$T=0.7\,\Tcs$, and the quartic-sector enhancement
\eqref{eq:B1anatomy} lifts the net factor partway toward unity. At
strong fields, panel (b), the character of the response changes
qualitatively: the singlet-only diode is cut off at its paramagnetic
scale, $B\simeq1.6\,\pi\Tcs$ at $T=0.7\,\Tcs$, whereas with an attractive
triplet channel the diode persists to fields several times larger,
peaking near the enhanced critical field. The trade of peak amplitude for
an extended operating window, together with the ridge structure of
Fig.~\ref{fig:phase}(b), constitutes the experimentally accessible
signature of the mechanism.

Two caveats delimit the plots. In a narrow strip below $\Tcs$, where the
Zeeman condensation-energy scale $b^2C/\tilde\rho^2$ competes with $t$,
the response becomes nonlinear in $\delta$ and the linear-response diode
theory does not apply; this strip is excluded from the figures. Near the
divergence of $B_c$ the triplet fraction is large and the GL truncation
degrades; the curves there are indicative.

%=====================================================================
%=====================================================================
%=====================================================================
%=====================================================================
%=====================================================================
\section{Summary and discussion}
\label{sec:discussion}

We have identified field-induced triplet pairing as a source of Lifshitz
invariants and of the intrinsic superconducting diode effect in Ising
superconductors with broken basal mirror symmetry, and we have derived
the complete theory of the effect, from the symmetry-constrained
two-component GL functional through the microscopic evaluation of all of
its coefficients to the diode efficiency across the phase diagram. The
main properties of the solution are as follows.

First, the mechanism has a kinematic origin and a correspondingly
universal structure. The Cooper-pair momentum and the in-plane field
enter every conversion vertex only through the covariant combination
$\bv=\Bv-\frac{\alpha_R}{2}(\qv\times\zhat)$, which locks the ratio of
the gradient and field couplings to $-\alpha_R/2$ at all orders. The
entire GL functional is generated by the single pair-breaking function
$C$ and its two derivatives, so the theory has no independent
phenomenological constants once the three microscopic scales are fixed.

Second, the anomalous sector has a two-path structure with opposite
interference patterns: destructive in the quadratic terms, where the
Lifshitz invariant \eqref{eq:A1anatomy} vanishes at $T=\Tct$, and
constructive in the quartic terms. At weak fields the net effect of an
attractive triplet channel is a nearly temperature-independent
suppression of the efficiency; at strong fields, where the spin-orbit
limiting of the triplet channel is quenched, the collective path
dominates and the diode acquires an extended field window terminating at
the triplet-enhanced critical field, with the maximal efficiency along
the adjacent ridge.

It is instructive to compare the triplet-mediated Lifshitz invariants
with the conventional ones. In the Rashba superconductor the exact
weak-coupling Lifshitz invariant \cite{Hasan2024} is built from the
frequency kernel
$\Delta_\so^2/[\omega_n^3(\omega_n^2+\Delta_\so^2)]$, which decomposes
identically as
$1/\omega_n^3-1/[\omega_n(\omega_n^2+\Delta_\so^2)]$. The second,
spin-orbit-limited kernel is precisely the propagator of the
field-induced triplet correlations of Sec.~\ref{sec:quadmicro}; upon
summation it produces the $C$-function part of the published
coefficients. The same two kernels appear as the intra- and interband
contributions in the Ising-Rashba model of Ref.~\cite{Bankier2025},
where the enhancement of the diode effect at strong Ising coupling is
carried by the kernel that is immune to the spin-orbit limiting. In this
sense the decomposition identifies the portion of the known,
band-splitting Lifshitz invariants that is already of triplet-correlation
origin at the level of virtual pairs; the present work promotes these
correlations to a dynamical order parameter and computes the additional,
collective-mode contributions that appear when the triplet channel is
attractive. The two families of invariants also differ parametrically.
The conventional parity-mixing triplet admixture of noncentrosymmetric
superconductors is of relative order $\Delta_\so/E_F$ and enters the
observables with the corresponding suppression; the field-induced
amplitude is of order $\Delta_Z/\Delta_\so$ and is insensitive to the
Fermi energy, exceeding the parity-mixing fraction already at fields
$B\gtrsim\Delta_\so^2/E_F$, which is small on the scale of the critical
fields of Ising superconductors. The overall factor of $\alpha_R$,
mandated by the mirror selection rule, is common to both families.

Several extensions are left for future work. The band-asymmetry
contributions to the singlet-sector baseline, which dominate the
low-field response at strong Ising coupling \cite{Bankier2025}, arise
from the density-of-states and velocity splitting of the bands and lie
outside the spin-sector quasiclassical formalism used here; a unified
treatment would combine both within a single gradient expansion.
Disorder, which is known to reshape the interplay of the singlet and
triplet channels \cite{Mockli2020,Ilic2022,Hasan2025}, should broaden the function
$C$ and with it every coefficient of the theory. The warping-induced
threefold anisotropy of the efficiency, derived in
Appendix~\ref{app:warping}, provides a symmetry-locked angular
fingerprint. Finally, the predicted correlation between the diode ridge
and the triplet-enhanced phase boundary suggests a concrete experimental
strategy: tracking the efficiency of a gated or substrate-asymmetrized
Ising superconductor as the in-plane field is swept toward the
paramagnetic limit would probe the hidden triplet channel and the
field-induced $s+if$ state through a transport measurement.

%=====================================================================
%=====================================================================
%=====================================================================
%=====================================================================
%=====================================================================
\begin{acknowledgments}
The work of A. L. was supported by NSF Grant No. DMR-2452658 and H. I. Romnes Faculty Fellowship provided by the University of Wisconsin-Madison Office of the Vice Chancellor for Research and Graduate Education with funding from the Wisconsin Alumni Research Foundation. 
M. K. acknowledges the financial support from the Israel Science Foundation, Grant No. 2665/20. 
The authors acknowledge the use of Claude (Anthropic) \cite{Claude2026} in the preparation of the manuscript, including, in particular, numerical calculations, graphics, and symbolic verification of analytical results. The authors conceived the project and checked and validated all results.
\end{acknowledgments}

%=====================================================================
\appendix

\section{Matsubara sums}
\label{app:sums}

All sums run over $\omega_n=\pi T(2n+1)$, $n\in\mathbb Z$, and
$y=\Delta/\pi T$ with $\Delta$ a generic energy. The basic identity
\eqref{eq:Cidentity} follows from partial fractions:
\begin{align}
\pi T\sum_{n}\frac{\Delta^2}{|\omega_n|(\omega_n^2+\Delta^2)}
&=2\pi T\sum_{n\ge0}\Big[\frac1{\omega_n}
-\frac{\omega_n}{\omega_n^2+\Delta^2}\Big]
\nonumber\\
&\hspace{-5em}=\sum_{n\ge0}\Big[\frac{1}{n+\frac12}
-\frac{1}{2}\frac{1}{n+\frac12+\frac{iy}2}
-\frac{1}{2}\frac{1}{n+\frac12-\frac{iy}2}\Big]
\nonumber\\
&\hspace{-5em}=\Ree\,\psi\Big(\tfrac12+\tfrac{iy}2\Big)-\psi\Big(\tfrac12\Big)
=C(y),
\end{align}
using the series
$\psi(z)=-\gamma_E+\sum_{n\ge0}[(n+1)^{-1}-(n+z)^{-1}]$. The
free-fermion sums are
\begin{equation}
\pi T\sum_n\frac{1}{|\omega_n|^3}=\frac{7\zeta(3)}{4\pi^2T^2},
\qquad
\pi T\sum_n\frac{1}{|\omega_n|^5}=\frac{31\zeta(5)}{16\pi^4T^4} .
\end{equation}
Differentiating the basic identity with respect to $\Delta^2$, using
$\partial\rho/\partial\Delta^2=\rho/2\Delta^2$,
\begin{align}
\pi T\sum_n\frac{1}{|\omega_n|(\omega_n^2+\Delta^2)}
&=\frac{C(y)}{\Delta^2},
\\
\pi T\sum_n\frac{|\omega_n|}{(\omega_n^2+\Delta^2)^2}
&=\frac{y\,C'(y)}{2\Delta^2},
\\
\pi T\sum_n\frac{2|\omega_n|(\omega_n^2-3\Delta^2)}
{(\omega_n^2+\Delta^2)^3}&=\frac{C''(y)}{(\pi T)^2} .
\label{eq:appCpp}
\end{align}
The last identity is obtained by combining the second derivative of the
basic identity with the free sums; it was verified numerically, 
as were all identities of this appendix. Closed expressions such as
$\pi T\sum_n(\omega_n^2+\Delta^2)^{-1}
=\tfrac{\pi}{2\Delta}\tanh\tfrac{\Delta}{2T}$ provide independent
checks.

The equivalence of the quartic coefficients \eqref{eq:betasC} with the
polygamma forms of Ref.~\cite{Mockli2019} rests on
\begin{equation}
\Imm\,\psi^{(1)}\Big(\tfrac12-\tfrac{i\rho}2\Big)=2C'(\rho),
\quad
\Ree\,\psi^{(2)}\Big(\tfrac12+\tfrac{i\rho}2\Big)=-4C''(\rho) ,
\end{equation}
which follow from Eq.~\eqref{eq:Cderivs}.

\section{Solution of the linearized equations and the kernels}
\label{app:kernels}

The free energy is generated by
\begin{align}
F=&-\frac12\sum\Delta^*V^{-1}\Delta
\nonumber\\
&+\frac1\beta\sum_{\kv,\omega_n}\sum_{l\ge1}\frac{(-2)^l}{2l}
\,\mathrm{tr}\big[G(\kv,\omega_n)\hat\Delta_{\kv}
G^{\rm T}(-\kv,-\omega_n)\hat\Delta^\dagger_{\kv}\big]^l ,
\label{eq:traceformula}
\end{align}
with $G^{-1}=i\omega_n-\xi_{\kv}-(\gv_{\kv}-\Bv)\cdot\bm\sigma$
\cite{Mockli2019}. At finite $\qv$ the substitution
\eqref{eq:replacements} applies, and the $l=1$ term is equivalent to the
linearized Eilenberger equations with the Doppler shift,
\begin{equation}
sf_0=\psi+i\bm f\cdot\bv,
\qquad
s\bm f=\dv+if_0\bv+\Delta_\so\hat\gamma\,(\zhat\times\bm f) ,
\label{eq:eilenbergerq}
\end{equation}
written for $\omega_n>0$, with $s=\omega_n+iu$ and the $f_z$ component
decoupled. Inverting the in-plane block,
\begin{equation}
\big[s-\Delta_\so\hat\gamma(\zhat\times)\big]^{-1}\bm x
=\frac{s\bm x+\Delta_\so\hat\gamma(\zhat\times\bm x)}
{s^2+\Delta_\so^2} ,
\end{equation}
and using $\bv\cdot(\zhat\times\bv)=0$, the singlet correlation is
\begin{equation}
f_0=\frac{\psi(s^2+\Delta_\so^2)
+i\hat\gamma s(\bv\cdot\etav)
-i\Delta_\so\zhat\cdot(\bv\times\etav)}{sW} ,
\label{eq:f0app}
\end{equation}
with $W=s^2+\Delta_\so^2+b^2$, and the triplet correlations follow by
back substitution,
\begin{equation}
\hat\gamma\bm f=\frac{s\etav+\Delta_\so(\zhat\times\etav)
+if_0[\hat\gamma s\bv+\Delta_\so(\zhat\times\bv)]}
{s^2+\Delta_\so^2} .
\label{eq:fapp}
\end{equation}
Substituting Eq.~\eqref{eq:f0app} into Eq.~\eqref{eq:fapp}, projecting
onto $\hat{\bv}$ and $\zhat\times\hat{\bv}$, and keeping the
$\hat\gamma$-even terms (the $\hat\gamma$-odd terms are the source of the
warping vertex, Appendix~\ref{app:warping}), the self-consistency
conditions of the two channels,
\begin{align}
\psi\ln\frac{T}{\Tcs}
&+\pi T\sum_n\Big(\frac{\psi}{|\omega_n|}
-\big\langle f_0\big\rangle\Big)=0,
\nonumber\\
\eta_i\ln\frac{T}{\Tct}
&+\pi T\sum_n\Big(\frac{\eta_i}{|\omega_n|}
-\big\langle\hat\gamma f_i\big\rangle\Big)=0 ,
\end{align}
assemble into the quadratic form with the kernels \eqref{eq:kernels}.
Two details deserve mention. The term
$\Delta_\so(\zhat\times\etav)/(s^2+\Delta_\so^2)$ in
Eq.~\eqref{eq:fapp} is odd under $\omega_n\to-\omega_n$ and cancels upon
frequency summation; it would otherwise generate an antisymmetric
bilinear that is forbidden in a real free energy. The conversion entry is
purely imaginary at the level of the complex order parameters, which is
the statement that the coupling has the $\Imm\{\psi^*\cdots\etav\}$ form
of Eq.~\eqref{eq:f2}, and becomes the real off-diagonal entry
$\alpha_{st}$ in the real-field convention of Sec.~\ref{sec:quadmicro}. The
zero-Doppler sums reproducing Eq.~\eqref{eq:MKmatrix} use the partial
fractions
\begin{equation}
\frac{b^2}{(\omega^2+\Delta_\so^2)(\omega^2+\tilde\rho^2)}
=\frac{1}{\omega^2+\Delta_\so^2}-\frac{1}{\omega^2+\tilde\rho^2} ,
\end{equation}
after which every sum is of the form of Appendix~\ref{app:sums}.

\section{Warping vertex}
\label{app:warping}

The $\hat\gamma$-odd part of Eq.~\eqref{eq:f0app} contributes to the
singlet self-consistency through the average
$\langle\hat\gamma\,h(u)\rangle_\theta$, which is nonzero only for the
harmonics of $h$ that overlap the threefold structure of $\hat\gamma$.
With $u=\tfrac12v_Fq\cos(\theta-\theta_q)$ the leading overlap is
\begin{equation}
\langle\hat\gamma u^3\rangle
=\frac{v_F^3q^3}{32}\,\bar w\,\cos3(\theta_q-\varphi_0),
\quad
\bar w=\begin{cases}2/\pi,&K\text{-model},\\[2pt]
1/\sqrt2,&\Gamma\text{-model},\end{cases}
\end{equation}
where $\varphi_0$ fixes the crystal axes. Expanding $1/W$ in
Eq.~\eqref{eq:f0app} to cubic order in $u$,
\begin{equation}
\Imm\frac1W\Big|_{u^3}
=u^3\Big[\frac{8\omega_n^3}{W_0^3}-\frac{4\omega_n}{W_0^2}\Big],
\qquad W_0=\omega_n^2+\tilde\rho^2 ,
\end{equation}
one obtains a coupling of the real part,
$F_2\supset2c_w(\qv)\Ree\{\psi^*(\bv\cdot\etav)\}$, with
\begin{equation}
c_w(\qv)=\frac{v_F^3q^3\bar w}{32}\cos3(\theta_q-\varphi_0)\;
\pi T\sum_n\Big[\frac{4\omega_n}{W_0^2}
-\frac{8\omega_n^3}{W_0^3}\Big] ,
\label{eq:cwapp}
\end{equation}
the remaining sum reducing to the $C$ family through the identities of
Appendix~\ref{app:sums}. The vertex couples the singlet to the doublet
projection along $\hat{\bv}$, orthogonal to the one selected by the
isotropic conversion, consistently with the symmetry analysis of
Sec.~\ref{sec:irreps}: the contraction $\bv\cdot\etav$ is odd under the
vertical mirrors and pairs with the equally odd threefold harmonic. Its
magnitude is suppressed by
$(v_Fq/\tilde\rho)^2\sim(T/\Delta_\so)^2$ relative to the isotropic
couplings. After elimination of the triplet it imprints a threefold
modulation on the diode efficiency as the current direction is rotated
with respect to the crystal axes, largest in the $\Gamma$-model. For the
$\Gamma$-model all kernels of Eq.~\eqref{eq:kernels} additionally acquire
$\hat\gamma^2$-weighted angular averages, which replace $C$ by the
corresponding weighted digamma averages without changing the structure of
the theory.

\section{Numerical procedure}
\label{app:numerics}

The exact evaluation proceeds as follows. For given
$(T,B,q)$ the entries \eqref{eq:entries} are computed by direct
summation, with the angular average on a uniform grid and the frequency
sum truncated adaptively (the difference form of the kernels converges as
$\omega_n^{-3}$). The condensation energy at fixed $q$ is obtained by
minimizing the quartic form over the two amplitudes, reduced to a
one-dimensional minimization over the ratio $r=\eta_c/\psi$ with the
amplitude optimized analytically. The supercurrent is evaluated by
high-order finite differences of $F(q)$, the window of $F<0$ is bracketed,
and the extrema of $J(q)$ are located on a grid and refined locally;
efficiencies are converged to five digits against grid refinement. Checks:
$\eSDE\to0$ linearity in $\alpha_R$ to
$0.3\%$ for $\delta\in[0.05,0.2]$ away from the near-$\Tcs$ strip; the
closed forms \eqref{eq:Astq}, \eqref{eq:xitriplet}, and
\eqref{eq:betasC} agree with direct summation to at least three
significant digits (seven for the conversion vertex); and the
perturbative efficiency assembled from the Taylor coefficients of
$\Aeff$ and $\Beff$ via Eq.~\eqref{eq:etamaster} reproduces the exact
one at the few-percent level near $\Tcs$.

\bibliography{refs}

\begin{thebibliography}{62}%
\makeatletter
\providecommand \@ifxundefined [1]{%
 \@ifx{#1\undefined}
}%
\providecommand \@ifnum [1]{%
 \ifnum #1\expandafter \@firstoftwo
 \else \expandafter \@secondoftwo
 \fi
}%
\providecommand \@ifx [1]{%
 \ifx #1\expandafter \@firstoftwo
 \else \expandafter \@secondoftwo
 \fi
}%
\providecommand \natexlab [1]{#1}%
\providecommand \enquote  [1]{``#1''}%
\providecommand \bibnamefont  [1]{#1}%
\providecommand \bibfnamefont [1]{#1}%
\providecommand \citenamefont [1]{#1}%
\providecommand \href@noop [0]{\@secondoftwo}%
\providecommand \href [0]{\begingroup \@sanitize@url \@href}%
\providecommand \@href[1]{\@@startlink{#1}\@@href}%
\providecommand \@@href[1]{\endgroup#1\@@endlink}%
\providecommand \@sanitize@url [0]{\catcode `\\12\catcode `\$12\catcode
  `\&12\catcode `\#12\catcode `\^12\catcode `\_12\catcode `\%12\relax}%
\providecommand \@@startlink[1]{}%
\providecommand \@@endlink[0]{}%
\providecommand \url  [0]{\begingroup\@sanitize@url \@url }%
\providecommand \@url [1]{\endgroup\@href {#1}{\urlprefix }}%
\providecommand \urlprefix  [0]{URL }%
\providecommand \Eprint [0]{\href }%
\providecommand \doibase [0]{http://dx.doi.org/}%
\providecommand \selectlanguage [0]{\@gobble}%
\providecommand \bibinfo  [0]{\@secondoftwo}%
\providecommand \bibfield  [0]{\@secondoftwo}%
\providecommand \translation [1]{[#1]}%
\providecommand \BibitemOpen [0]{}%
\providecommand \bibitemStop [0]{}%
\providecommand \bibitemNoStop [0]{.\EOS\space}%
\providecommand \EOS [0]{\spacefactor3000\relax}%
\providecommand \BibitemShut  [1]{\csname bibitem#1\endcsname}%
\let\auto@bib@innerbib\@empty
%</preamble>
\bibitem [{\citenamefont {Ando}\ \emph {et~al.}(2020)\citenamefont {Ando},
  \citenamefont {Miyasaka}, \citenamefont {Li}, \citenamefont {Ishizuka},
  \citenamefont {Arakawa}, \citenamefont {Shiota}, \citenamefont {Moriyama},
  \citenamefont {Yanase},\ and\ \citenamefont {Ono}}]{Ando2020}%
  \BibitemOpen
  \bibfield  {author} {\bibinfo {author} {\bibfnamefont {F.}~\bibnamefont
  {Ando}}, \bibinfo {author} {\bibfnamefont {Y.}~\bibnamefont {Miyasaka}},
  \bibinfo {author} {\bibfnamefont {T.}~\bibnamefont {Li}}, \bibinfo {author}
  {\bibfnamefont {J.}~\bibnamefont {Ishizuka}}, \bibinfo {author}
  {\bibfnamefont {T.}~\bibnamefont {Arakawa}}, \bibinfo {author} {\bibfnamefont
  {Y.}~\bibnamefont {Shiota}}, \bibinfo {author} {\bibfnamefont
  {T.}~\bibnamefont {Moriyama}}, \bibinfo {author} {\bibfnamefont
  {Y.}~\bibnamefont {Yanase}}, \ and\ \bibinfo {author} {\bibfnamefont
  {T.}~\bibnamefont {Ono}},\ }\bibfield  {title} {\enquote {\bibinfo {title}
  {Observation of superconducting diode effect},}\ }\href@noop {} {\bibfield
  {journal} {\bibinfo  {journal} {Nature (London)}\ }\textbf {\bibinfo {volume}
  {584}},\ \bibinfo {pages} {373} (\bibinfo {year} {2020})}\BibitemShut
  {NoStop}%
\bibitem [{\citenamefont {Nadeem}\ \emph {et~al.}(2023)\citenamefont {Nadeem},
  \citenamefont {Fuhrer},\ and\ \citenamefont {Wang}}]{Nadeem2023}%
  \BibitemOpen
  \bibfield  {author} {\bibinfo {author} {\bibfnamefont {M.}~\bibnamefont
  {Nadeem}}, \bibinfo {author} {\bibfnamefont {M.~S.}\ \bibnamefont {Fuhrer}},
  \ and\ \bibinfo {author} {\bibfnamefont {X.}~\bibnamefont {Wang}},\
  }\bibfield  {title} {\enquote {\bibinfo {title} {The superconducting diode
  effect},}\ }\href@noop {} {\bibfield  {journal} {\bibinfo  {journal} {Nat.
  Rev. Phys.}\ }\textbf {\bibinfo {volume} {5}},\ \bibinfo {pages} {558}
  (\bibinfo {year} {2023})}\BibitemShut {NoStop}%
\bibitem [{\citenamefont {Jiang}\ and\ \citenamefont {Hu}(2022)}]{Jiang2022}%
  \BibitemOpen
  \bibfield  {author} {\bibinfo {author} {\bibfnamefont {K.}~\bibnamefont
  {Jiang}}\ and\ \bibinfo {author} {\bibfnamefont {J.}~\bibnamefont {Hu}},\
  }\bibfield  {title} {\enquote {\bibinfo {title} {Superconducting diode
  effects},}\ }\href@noop {} {\bibfield  {journal} {\bibinfo  {journal} {Nat.
  Phys.}\ }\textbf {\bibinfo {volume} {18}},\ \bibinfo {pages} {1145} (\bibinfo
  {year} {2022})}\BibitemShut {NoStop}%
\bibitem [{\citenamefont {Ma}\ \emph {et~al.}(2025)\citenamefont {Ma},
  \citenamefont {Zhan},\ and\ \citenamefont {Lin}}]{Ma2025}%
  \BibitemOpen
  \bibfield  {author} {\bibinfo {author} {\bibfnamefont {J.}~\bibnamefont
  {Ma}}, \bibinfo {author} {\bibfnamefont {R.}~\bibnamefont {Zhan}}, \ and\
  \bibinfo {author} {\bibfnamefont {X.}~\bibnamefont {Lin}},\ }\bibfield
  {title} {\enquote {\bibinfo {title} {Superconducting diode effects:
  Mechanisms, materials and applications},}\ }\href@noop {} {\bibfield
  {journal} {\bibinfo  {journal} {Adv. Phys. Res.}\ }\textbf {\bibinfo {volume}
  {4}},\ \bibinfo {pages} {2400180} (\bibinfo {year} {2025})}\BibitemShut
  {NoStop}%
\bibitem [{\citenamefont {Shaffer}\ and\ \citenamefont
  {Levchenko}(2025)}]{ShafferLevchenko2025}%
  \BibitemOpen
  \bibfield  {author} {\bibinfo {author} {\bibfnamefont {Daniel}\ \bibnamefont
  {Shaffer}}\ and\ \bibinfo {author} {\bibfnamefont {Alex}\ \bibnamefont
  {Levchenko}},\ }\href@noop {} {\enquote {\bibinfo {title} {Theories of
  superconducting diode effects},}\ } (\bibinfo {year} {2025}),\ \Eprint
  {http://arxiv.org/abs/2510.25864} {arXiv:2510.25864 [cond-mat.supr-con]}
  \BibitemShut {NoStop}%
\bibitem [{\citenamefont {Baumgartner}\ \emph {et~al.}(2022)\citenamefont
  {Baumgartner}, \citenamefont {Fuchs}, \citenamefont {Costa}, \citenamefont
  {Reinhardt}, \citenamefont {Gronin}, \citenamefont {Gardner}, \citenamefont
  {Lindemann}, \citenamefont {Manfra}, \citenamefont {Faria~Junior},
  \citenamefont {Kochan}, \citenamefont {Fabian}, \citenamefont {Paradiso},\
  and\ \citenamefont {Strunk}}]{Baumgartner2022}%
  \BibitemOpen
  \bibfield  {author} {\bibinfo {author} {\bibfnamefont {C.}~\bibnamefont
  {Baumgartner}}, \bibinfo {author} {\bibfnamefont {L.}~\bibnamefont {Fuchs}},
  \bibinfo {author} {\bibfnamefont {A.}~\bibnamefont {Costa}}, \bibinfo
  {author} {\bibfnamefont {S.}~\bibnamefont {Reinhardt}}, \bibinfo {author}
  {\bibfnamefont {S.}~\bibnamefont {Gronin}}, \bibinfo {author} {\bibfnamefont
  {G.~C.}\ \bibnamefont {Gardner}}, \bibinfo {author} {\bibfnamefont
  {T.}~\bibnamefont {Lindemann}}, \bibinfo {author} {\bibfnamefont {M.~J.}\
  \bibnamefont {Manfra}}, \bibinfo {author} {\bibfnamefont {P.~E.}\
  \bibnamefont {Faria~Junior}}, \bibinfo {author} {\bibfnamefont
  {D.}~\bibnamefont {Kochan}}, \bibinfo {author} {\bibfnamefont
  {J.}~\bibnamefont {Fabian}}, \bibinfo {author} {\bibfnamefont
  {N.}~\bibnamefont {Paradiso}}, \ and\ \bibinfo {author} {\bibfnamefont
  {C.}~\bibnamefont {Strunk}},\ }\bibfield  {title} {\enquote {\bibinfo {title}
  {Supercurrent rectification and magnetochiral effects in symmetric
  {Josephson} junctions},}\ }\href@noop {} {\bibfield  {journal} {\bibinfo
  {journal} {Nat. Nanotechnol.}\ }\textbf {\bibinfo {volume} {17}},\ \bibinfo
  {pages} {39} (\bibinfo {year} {2022})}\BibitemShut {NoStop}%
\bibitem [{\citenamefont {Wu}\ \emph {et~al.}(2022)\citenamefont {Wu},
  \citenamefont {Wang}, \citenamefont {Xu}, \citenamefont {Sivakumar},
  \citenamefont {Pasco}, \citenamefont {Filippozzi}, \citenamefont {Parkin},
  \citenamefont {Zeng}, \citenamefont {McQueen},\ and\ \citenamefont
  {Ali}}]{Wu2022}%
  \BibitemOpen
  \bibfield  {author} {\bibinfo {author} {\bibfnamefont {H.}~\bibnamefont
  {Wu}}, \bibinfo {author} {\bibfnamefont {Y.}~\bibnamefont {Wang}}, \bibinfo
  {author} {\bibfnamefont {Y.}~\bibnamefont {Xu}}, \bibinfo {author}
  {\bibfnamefont {P.~K.}\ \bibnamefont {Sivakumar}}, \bibinfo {author}
  {\bibfnamefont {C.}~\bibnamefont {Pasco}}, \bibinfo {author} {\bibfnamefont
  {U.}~\bibnamefont {Filippozzi}}, \bibinfo {author} {\bibfnamefont {S.~S.~P.}\
  \bibnamefont {Parkin}}, \bibinfo {author} {\bibfnamefont {Y.-J.}\
  \bibnamefont {Zeng}}, \bibinfo {author} {\bibfnamefont {T.}~\bibnamefont
  {McQueen}}, \ and\ \bibinfo {author} {\bibfnamefont {M.~N.}\ \bibnamefont
  {Ali}},\ }\bibfield  {title} {\enquote {\bibinfo {title} {The field-free
  {Josephson} diode in a van der {Waals} heterostructure},}\ }\href@noop {}
  {\bibfield  {journal} {\bibinfo  {journal} {Nature (London)}\ }\textbf
  {\bibinfo {volume} {604}},\ \bibinfo {pages} {653} (\bibinfo {year}
  {2022})}\BibitemShut {NoStop}%
\bibitem [{\citenamefont {Lyu}\ \emph {et~al.}(2021)\citenamefont {Lyu},
  \citenamefont {Jiang}, \citenamefont {Wang}, \citenamefont {Xiao},
  \citenamefont {Dong}, \citenamefont {Chen}, \citenamefont {Milo{\v
  s}evi{\'c}}, \citenamefont {Wang}, \citenamefont {Divan}, \citenamefont
  {Pearson}, \citenamefont {Wu}, \citenamefont {Peeters},\ and\ \citenamefont
  {Kwok}}]{Lyu2021}%
  \BibitemOpen
  \bibfield  {author} {\bibinfo {author} {\bibfnamefont {Y.-Y.}\ \bibnamefont
  {Lyu}}, \bibinfo {author} {\bibfnamefont {J.}~\bibnamefont {Jiang}}, \bibinfo
  {author} {\bibfnamefont {Y.-L.}\ \bibnamefont {Wang}}, \bibinfo {author}
  {\bibfnamefont {Z.-L.}\ \bibnamefont {Xiao}}, \bibinfo {author}
  {\bibfnamefont {S.}~\bibnamefont {Dong}}, \bibinfo {author} {\bibfnamefont
  {Q.-H.}\ \bibnamefont {Chen}}, \bibinfo {author} {\bibfnamefont {M.~V.}\
  \bibnamefont {Milo{\v s}evi{\'c}}}, \bibinfo {author} {\bibfnamefont
  {H.}~\bibnamefont {Wang}}, \bibinfo {author} {\bibfnamefont {R.}~\bibnamefont
  {Divan}}, \bibinfo {author} {\bibfnamefont {J.~E.}\ \bibnamefont {Pearson}},
  \bibinfo {author} {\bibfnamefont {P.}~\bibnamefont {Wu}}, \bibinfo {author}
  {\bibfnamefont {F.~M.}\ \bibnamefont {Peeters}}, \ and\ \bibinfo {author}
  {\bibfnamefont {W.-K.}\ \bibnamefont {Kwok}},\ }\bibfield  {title} {\enquote
  {\bibinfo {title} {Superconducting diode effect via conformal-mapped
  nanoholes},}\ }\href@noop {} {\bibfield  {journal} {\bibinfo  {journal} {Nat.
  Commun.}\ }\textbf {\bibinfo {volume} {12}},\ \bibinfo {pages} {2703}
  (\bibinfo {year} {2021})}\BibitemShut {NoStop}%
\bibitem [{\citenamefont {Vodolazov}\ and\ \citenamefont
  {Peeters}(2005)}]{Vodolazov2005}%
  \BibitemOpen
  \bibfield  {author} {\bibinfo {author} {\bibfnamefont {D.~Y.}\ \bibnamefont
  {Vodolazov}}\ and\ \bibinfo {author} {\bibfnamefont {F.~M.}\ \bibnamefont
  {Peeters}},\ }\bibfield  {title} {\enquote {\bibinfo {title} {Superconducting
  rectifier based on the asymmetric surface barrier effect},}\ }\href@noop {}
  {\bibfield  {journal} {\bibinfo  {journal} {Phys. Rev. B}\ }\textbf {\bibinfo
  {volume} {72}},\ \bibinfo {pages} {172508} (\bibinfo {year}
  {2005})}\BibitemShut {NoStop}%
\bibitem [{\citenamefont {Hou}\ \emph {et~al.}(2023)\citenamefont {Hou},
  \citenamefont {Nichele}, \citenamefont {Chi}, \citenamefont {Lodesani},
  \citenamefont {Wu}, \citenamefont {Ritter}, \citenamefont {Haxell},
  \citenamefont {Davydova}, \citenamefont {Ili{\'c}}, \citenamefont
  {Glezakou-Elbert}, \citenamefont {Varambally}, \citenamefont {Bergeret},
  \citenamefont {Kamra}, \citenamefont {Fu}, \citenamefont {Lee},\ and\
  \citenamefont {Moodera}}]{Hou2023}%
  \BibitemOpen
  \bibfield  {author} {\bibinfo {author} {\bibfnamefont {Y.}~\bibnamefont
  {Hou}}, \bibinfo {author} {\bibfnamefont {F.}~\bibnamefont {Nichele}},
  \bibinfo {author} {\bibfnamefont {H.}~\bibnamefont {Chi}}, \bibinfo {author}
  {\bibfnamefont {A.}~\bibnamefont {Lodesani}}, \bibinfo {author}
  {\bibfnamefont {Y.}~\bibnamefont {Wu}}, \bibinfo {author} {\bibfnamefont
  {M.~F.}\ \bibnamefont {Ritter}}, \bibinfo {author} {\bibfnamefont {D.~Z.}\
  \bibnamefont {Haxell}}, \bibinfo {author} {\bibfnamefont {M.}~\bibnamefont
  {Davydova}}, \bibinfo {author} {\bibfnamefont {S.}~\bibnamefont {Ili{\'c}}},
  \bibinfo {author} {\bibfnamefont {O.}~\bibnamefont {Glezakou-Elbert}},
  \bibinfo {author} {\bibfnamefont {A.}~\bibnamefont {Varambally}}, \bibinfo
  {author} {\bibfnamefont {F.~S.}\ \bibnamefont {Bergeret}}, \bibinfo {author}
  {\bibfnamefont {A.}~\bibnamefont {Kamra}}, \bibinfo {author} {\bibfnamefont
  {L.}~\bibnamefont {Fu}}, \bibinfo {author} {\bibfnamefont {P.~A.}\
  \bibnamefont {Lee}}, \ and\ \bibinfo {author} {\bibfnamefont {J.~S.}\
  \bibnamefont {Moodera}},\ }\bibfield  {title} {\enquote {\bibinfo {title}
  {Ubiquitous superconducting diode effect in superconductor thin films},}\
  }\href@noop {} {\bibfield  {journal} {\bibinfo  {journal} {Phys. Rev. Lett.}\
  }\textbf {\bibinfo {volume} {131}},\ \bibinfo {pages} {027001} (\bibinfo
  {year} {2023})}\BibitemShut {NoStop}%
\bibitem [{\citenamefont {Sundaresh}\ \emph {et~al.}(2023)\citenamefont
  {Sundaresh}, \citenamefont {V{\"a}yrynen}, \citenamefont {Lyanda-Geller},\
  and\ \citenamefont {Rokhinson}}]{Sundaresh2023}%
  \BibitemOpen
  \bibfield  {author} {\bibinfo {author} {\bibfnamefont {A.}~\bibnamefont
  {Sundaresh}}, \bibinfo {author} {\bibfnamefont {J.~I.}\ \bibnamefont
  {V{\"a}yrynen}}, \bibinfo {author} {\bibfnamefont {Y.}~\bibnamefont
  {Lyanda-Geller}}, \ and\ \bibinfo {author} {\bibfnamefont {L.~P.}\
  \bibnamefont {Rokhinson}},\ }\bibfield  {title} {\enquote {\bibinfo {title}
  {Diamagnetic mechanism of critical current nonreciprocity in multilayered
  superconductors},}\ }\href@noop {} {\bibfield  {journal} {\bibinfo  {journal}
  {Nat. Commun.}\ }\textbf {\bibinfo {volume} {14}},\ \bibinfo {pages} {1628}
  (\bibinfo {year} {2023})}\BibitemShut {NoStop}%
\bibitem [{\citenamefont {Gutfreund}\ \emph {et~al.}(2023)\citenamefont
  {Gutfreund}, \citenamefont {Matsuki}, \citenamefont {Plastovets},
  \citenamefont {Noah}, \citenamefont {Gorzawski}, \citenamefont {Fridman},
  \citenamefont {Yang}, \citenamefont {Buzdin}, \citenamefont {Millo},
  \citenamefont {Robinson},\ and\ \citenamefont {Anahory}}]{Gutfreund2023}%
  \BibitemOpen
  \bibfield  {author} {\bibinfo {author} {\bibfnamefont {A.}~\bibnamefont
  {Gutfreund}}, \bibinfo {author} {\bibfnamefont {H.}~\bibnamefont {Matsuki}},
  \bibinfo {author} {\bibfnamefont {V.}~\bibnamefont {Plastovets}}, \bibinfo
  {author} {\bibfnamefont {A.}~\bibnamefont {Noah}}, \bibinfo {author}
  {\bibfnamefont {L.}~\bibnamefont {Gorzawski}}, \bibinfo {author}
  {\bibfnamefont {N.}~\bibnamefont {Fridman}}, \bibinfo {author} {\bibfnamefont
  {G.}~\bibnamefont {Yang}}, \bibinfo {author} {\bibfnamefont {A.}~\bibnamefont
  {Buzdin}}, \bibinfo {author} {\bibfnamefont {O.}~\bibnamefont {Millo}},
  \bibinfo {author} {\bibfnamefont {J.~W.~A.}\ \bibnamefont {Robinson}}, \ and\
  \bibinfo {author} {\bibfnamefont {Y.}~\bibnamefont {Anahory}},\ }\bibfield
  {title} {\enquote {\bibinfo {title} {Direct observation of a superconducting
  vortex diode},}\ }\href@noop {} {\bibfield  {journal} {\bibinfo  {journal}
  {Nat. Commun.}\ }\textbf {\bibinfo {volume} {14}},\ \bibinfo {pages} {1630}
  (\bibinfo {year} {2023})}\BibitemShut {NoStop}%
\bibitem [{\citenamefont {Edelstein}(1996)}]{Edelstein1996}%
  \BibitemOpen
  \bibfield  {author} {\bibinfo {author} {\bibfnamefont {V.~M.}\ \bibnamefont
  {Edelstein}},\ }\bibfield  {title} {\enquote {\bibinfo {title} {The
  {Ginzburg-Landau} equation for superconductors of polar symmetry},}\
  }\href@noop {} {\bibfield  {journal} {\bibinfo  {journal} {J. Phys.: Condens.
  Matter}\ }\textbf {\bibinfo {volume} {8}},\ \bibinfo {pages} {339} (\bibinfo
  {year} {1996})}\BibitemShut {NoStop}%
\bibitem [{\citenamefont {Daido}\ \emph {et~al.}(2022)\citenamefont {Daido},
  \citenamefont {Ikeda},\ and\ \citenamefont {Yanase}}]{Daido2022}%
  \BibitemOpen
  \bibfield  {author} {\bibinfo {author} {\bibfnamefont {A.}~\bibnamefont
  {Daido}}, \bibinfo {author} {\bibfnamefont {Y.}~\bibnamefont {Ikeda}}, \ and\
  \bibinfo {author} {\bibfnamefont {Y.}~\bibnamefont {Yanase}},\ }\bibfield
  {title} {\enquote {\bibinfo {title} {Intrinsic superconducting diode
  effect},}\ }\href@noop {} {\bibfield  {journal} {\bibinfo  {journal} {Phys.
  Rev. Lett.}\ }\textbf {\bibinfo {volume} {128}},\ \bibinfo {pages} {037001}
  (\bibinfo {year} {2022})}\BibitemShut {NoStop}%
\bibitem [{\citenamefont {Yuan}\ and\ \citenamefont {Fu}(2022)}]{Yuan2022}%
  \BibitemOpen
  \bibfield  {author} {\bibinfo {author} {\bibfnamefont {N.~F.~Q.}\
  \bibnamefont {Yuan}}\ and\ \bibinfo {author} {\bibfnamefont {L.}~\bibnamefont
  {Fu}},\ }\bibfield  {title} {\enquote {\bibinfo {title} {Supercurrent diode
  effect and finite-momentum superconductors},}\ }\href@noop {} {\bibfield
  {journal} {\bibinfo  {journal} {Proc. Natl. Acad. Sci. USA}\ }\textbf
  {\bibinfo {volume} {119}},\ \bibinfo {pages} {e2119548119} (\bibinfo {year}
  {2022})}\BibitemShut {NoStop}%
\bibitem [{\citenamefont {He}\ \emph {et~al.}(2022)\citenamefont {He},
  \citenamefont {Tanaka},\ and\ \citenamefont {Nagaosa}}]{He2022}%
  \BibitemOpen
  \bibfield  {author} {\bibinfo {author} {\bibfnamefont {J.~J.}\ \bibnamefont
  {He}}, \bibinfo {author} {\bibfnamefont {Y.}~\bibnamefont {Tanaka}}, \ and\
  \bibinfo {author} {\bibfnamefont {N.}~\bibnamefont {Nagaosa}},\ }\bibfield
  {title} {\enquote {\bibinfo {title} {A phenomenological theory of
  superconductor diodes},}\ }\href@noop {} {\bibfield  {journal} {\bibinfo
  {journal} {New J. Phys.}\ }\textbf {\bibinfo {volume} {24}},\ \bibinfo
  {pages} {053014} (\bibinfo {year} {2022})}\BibitemShut {NoStop}%
\bibitem [{\citenamefont {Ili{\'c}}\ and\ \citenamefont
  {Bergeret}(2022)}]{Ilic2022}%
  \BibitemOpen
  \bibfield  {author} {\bibinfo {author} {\bibfnamefont {S.}~\bibnamefont
  {Ili{\'c}}}\ and\ \bibinfo {author} {\bibfnamefont {F.~S.}\ \bibnamefont
  {Bergeret}},\ }\bibfield  {title} {\enquote {\bibinfo {title} {Theory of the
  supercurrent diode effect in {Rashba} superconductors with arbitrary
  disorder},}\ }\href@noop {} {\bibfield  {journal} {\bibinfo  {journal} {Phys.
  Rev. Lett.}\ }\textbf {\bibinfo {volume} {128}},\ \bibinfo {pages} {177001}
  (\bibinfo {year} {2022})}\BibitemShut {NoStop}%
\bibitem [{\citenamefont {Hasan}\ \emph {et~al.}(2024)\citenamefont {Hasan},
  \citenamefont {Shaffer}, \citenamefont {Khodas},\ and\ \citenamefont
  {Levchenko}}]{Hasan2024}%
  \BibitemOpen
  \bibfield  {author} {\bibinfo {author} {\bibfnamefont {J.}~\bibnamefont
  {Hasan}}, \bibinfo {author} {\bibfnamefont {D.}~\bibnamefont {Shaffer}},
  \bibinfo {author} {\bibfnamefont {M.}~\bibnamefont {Khodas}}, \ and\ \bibinfo
  {author} {\bibfnamefont {A.}~\bibnamefont {Levchenko}},\ }\bibfield  {title}
  {\enquote {\bibinfo {title} {Supercurrent diode effect in helical
  superconductors},}\ }\href@noop {} {\bibfield  {journal} {\bibinfo  {journal}
  {Phys. Rev. B}\ }\textbf {\bibinfo {volume} {110}},\ \bibinfo {pages}
  {024508} (\bibinfo {year} {2024})}\BibitemShut {NoStop}%
\bibitem [{\citenamefont {Hasan}\ \emph {et~al.}(2025)\citenamefont {Hasan},
  \citenamefont {Shaffer}, \citenamefont {Khodas},\ and\ \citenamefont
  {Levchenko}}]{Hasan2025}%
  \BibitemOpen
  \bibfield  {author} {\bibinfo {author} {\bibfnamefont {Jaglul}\ \bibnamefont
  {Hasan}}, \bibinfo {author} {\bibfnamefont {Daniel}\ \bibnamefont {Shaffer}},
  \bibinfo {author} {\bibfnamefont {Maxim}\ \bibnamefont {Khodas}}, \ and\
  \bibinfo {author} {\bibfnamefont {Alex}\ \bibnamefont {Levchenko}},\
  }\bibfield  {title} {\enquote {\bibinfo {title} {Superconducting diode
  efficiency from singlet-triplet mixing in disordered systems},}\ }\href
  {\doibase 10.1103/PhysRevB.111.174514} {\bibfield  {journal} {\bibinfo
  {journal} {Phys. Rev. B}\ }\textbf {\bibinfo {volume} {111}},\ \bibinfo
  {pages} {174514} (\bibinfo {year} {2025})}\BibitemShut {NoStop}%
\bibitem [{\citenamefont {Bankier}\ \emph {et~al.}(2025)\citenamefont
  {Bankier}, \citenamefont {Attias}, \citenamefont {Levchenko},\ and\
  \citenamefont {Khodas}}]{Bankier2025}%
  \BibitemOpen
  \bibfield  {author} {\bibinfo {author} {\bibfnamefont {I.}~\bibnamefont
  {Bankier}}, \bibinfo {author} {\bibfnamefont {L.}~\bibnamefont {Attias}},
  \bibinfo {author} {\bibfnamefont {A.}~\bibnamefont {Levchenko}}, \ and\
  \bibinfo {author} {\bibfnamefont {M.}~\bibnamefont {Khodas}},\ }\bibfield
  {title} {\enquote {\bibinfo {title} {Superconducting diode effect in {Ising}
  superconductors},}\ }\href@noop {} {\bibfield  {journal} {\bibinfo  {journal}
  {Phys. Rev. B}\ }\textbf {\bibinfo {volume} {111}},\ \bibinfo {pages}
  {L180505} (\bibinfo {year} {2025})}\BibitemShut {NoStop}%
\bibitem [{\citenamefont {Wakatsuki}\ and\ \citenamefont
  {Nagaosa}(2018)}]{Wakatsuki2018}%
  \BibitemOpen
  \bibfield  {author} {\bibinfo {author} {\bibfnamefont {R.}~\bibnamefont
  {Wakatsuki}}\ and\ \bibinfo {author} {\bibfnamefont {N.}~\bibnamefont
  {Nagaosa}},\ }\bibfield  {title} {\enquote {\bibinfo {title} {Nonreciprocal
  current in noncentrosymmetric {Rashba} superconductors},}\ }\href@noop {}
  {\bibfield  {journal} {\bibinfo  {journal} {Phys. Rev. Lett.}\ }\textbf
  {\bibinfo {volume} {121}},\ \bibinfo {pages} {026601} (\bibinfo {year}
  {2018})}\BibitemShut {NoStop}%
\bibitem [{\citenamefont {Hoshino}\ \emph {et~al.}(2018)\citenamefont
  {Hoshino}, \citenamefont {Wakatsuki}, \citenamefont {Hamamoto},\ and\
  \citenamefont {Nagaosa}}]{Hoshino2018}%
  \BibitemOpen
  \bibfield  {author} {\bibinfo {author} {\bibfnamefont {S.}~\bibnamefont
  {Hoshino}}, \bibinfo {author} {\bibfnamefont {R.}~\bibnamefont {Wakatsuki}},
  \bibinfo {author} {\bibfnamefont {K.}~\bibnamefont {Hamamoto}}, \ and\
  \bibinfo {author} {\bibfnamefont {N.}~\bibnamefont {Nagaosa}},\ }\bibfield
  {title} {\enquote {\bibinfo {title} {Nonreciprocal charge transport in
  two-dimensional noncentrosymmetric superconductors},}\ }\href@noop {}
  {\bibfield  {journal} {\bibinfo  {journal} {Phys. Rev. B}\ }\textbf {\bibinfo
  {volume} {98}},\ \bibinfo {pages} {054510} (\bibinfo {year}
  {2018})}\BibitemShut {NoStop}%
\bibitem [{\citenamefont {Zinkl}\ \emph {et~al.}(2022)\citenamefont {Zinkl},
  \citenamefont {Hamamoto},\ and\ \citenamefont {Sigrist}}]{Zinkl2022}%
  \BibitemOpen
  \bibfield  {author} {\bibinfo {author} {\bibfnamefont {B.}~\bibnamefont
  {Zinkl}}, \bibinfo {author} {\bibfnamefont {K.}~\bibnamefont {Hamamoto}}, \
  and\ \bibinfo {author} {\bibfnamefont {M.}~\bibnamefont {Sigrist}},\
  }\bibfield  {title} {\enquote {\bibinfo {title} {Symmetry conditions for the
  superconducting diode effect in chiral superconductors},}\ }\href@noop {}
  {\bibfield  {journal} {\bibinfo  {journal} {Phys. Rev. Res.}\ }\textbf
  {\bibinfo {volume} {4}},\ \bibinfo {pages} {033167} (\bibinfo {year}
  {2022})}\BibitemShut {NoStop}%
\bibitem [{\citenamefont {Daido}\ and\ \citenamefont
  {Yanase}(2022)}]{Daido2022b}%
  \BibitemOpen
  \bibfield  {author} {\bibinfo {author} {\bibfnamefont {A.}~\bibnamefont
  {Daido}}\ and\ \bibinfo {author} {\bibfnamefont {Y.}~\bibnamefont {Yanase}},\
  }\bibfield  {title} {\enquote {\bibinfo {title} {Superconducting diode effect
  and nonreciprocal transition lines},}\ }\href@noop {} {\bibfield  {journal}
  {\bibinfo  {journal} {Phys. Rev. B}\ }\textbf {\bibinfo {volume} {106}},\
  \bibinfo {pages} {205206} (\bibinfo {year} {2022})}\BibitemShut {NoStop}%
\bibitem [{\citenamefont {Mineev}\ and\ \citenamefont
  {Samokhin}(1994)}]{Mineev1994}%
  \BibitemOpen
  \bibfield  {author} {\bibinfo {author} {\bibfnamefont {V.~P.}\ \bibnamefont
  {Mineev}}\ and\ \bibinfo {author} {\bibfnamefont {K.~V.}\ \bibnamefont
  {Samokhin}},\ }\bibfield  {title} {\enquote {\bibinfo {title} {Helical phases
  in superconductors},}\ }\href@noop {} {\bibfield  {journal} {\bibinfo
  {journal} {Zh. Eksp. Teor. Fiz.}\ }\textbf {\bibinfo {volume} {105}},\
  \bibinfo {pages} {747} (\bibinfo {year} {1994})},\ \bibinfo {note} {[Sov.
  Phys. JETP {\bf 78}, 401 (1994)]}\BibitemShut {NoStop}%
\bibitem [{\citenamefont {Agterberg}(2003)}]{Agterberg2003}%
  \BibitemOpen
  \bibfield  {author} {\bibinfo {author} {\bibfnamefont {D.~F.}\ \bibnamefont
  {Agterberg}},\ }\bibfield  {title} {\enquote {\bibinfo {title} {Novel
  magnetic field effects in unconventional superconductors},}\ }\href@noop {}
  {\bibfield  {journal} {\bibinfo  {journal} {Physica C}\ }\textbf {\bibinfo
  {volume} {387}},\ \bibinfo {pages} {13} (\bibinfo {year} {2003})}\BibitemShut
  {NoStop}%
\bibitem [{\citenamefont {Kaur}\ \emph {et~al.}(2005)\citenamefont {Kaur},
  \citenamefont {Agterberg},\ and\ \citenamefont {Sigrist}}]{Kaur2005}%
  \BibitemOpen
  \bibfield  {author} {\bibinfo {author} {\bibfnamefont {R.~P.}\ \bibnamefont
  {Kaur}}, \bibinfo {author} {\bibfnamefont {D.~F.}\ \bibnamefont {Agterberg}},
  \ and\ \bibinfo {author} {\bibfnamefont {M.}~\bibnamefont {Sigrist}},\
  }\bibfield  {title} {\enquote {\bibinfo {title} {Helical vortex phase in the
  noncentrosymmetric {CePt$_3$Si}},}\ }\href@noop {} {\bibfield  {journal}
  {\bibinfo  {journal} {Phys. Rev. Lett.}\ }\textbf {\bibinfo {volume} {94}},\
  \bibinfo {pages} {137002} (\bibinfo {year} {2005})}\BibitemShut {NoStop}%
\bibitem [{\citenamefont {Dimitrova}\ and\ \citenamefont
  {Feigel'man}(2007)}]{Dimitrova2007}%
  \BibitemOpen
  \bibfield  {author} {\bibinfo {author} {\bibfnamefont {O.}~\bibnamefont
  {Dimitrova}}\ and\ \bibinfo {author} {\bibfnamefont {M.~V.}\ \bibnamefont
  {Feigel'man}},\ }\bibfield  {title} {\enquote {\bibinfo {title} {Theory of a
  two-dimensional superconductor with broken inversion symmetry},}\ }\href@noop
  {} {\bibfield  {journal} {\bibinfo  {journal} {Phys. Rev. B}\ }\textbf
  {\bibinfo {volume} {76}},\ \bibinfo {pages} {014522} (\bibinfo {year}
  {2007})}\BibitemShut {NoStop}%
\bibitem [{\citenamefont {Mineev}\ and\ \citenamefont
  {Samokhin}(2008)}]{MineevSamokhin2008}%
  \BibitemOpen
  \bibfield  {author} {\bibinfo {author} {\bibfnamefont {V.~P.}\ \bibnamefont
  {Mineev}}\ and\ \bibinfo {author} {\bibfnamefont {K.~V.}\ \bibnamefont
  {Samokhin}},\ }\bibfield  {title} {\enquote {\bibinfo {title} {Nonuniform
  states in noncentrosymmetric superconductors: Derivation of {Lifshitz}
  invariants from microscopic theory},}\ }\href@noop {} {\bibfield  {journal}
  {\bibinfo  {journal} {Phys. Rev. B}\ }\textbf {\bibinfo {volume} {78}},\
  \bibinfo {pages} {144503} (\bibinfo {year} {2008})}\BibitemShut {NoStop}%
\bibitem [{\citenamefont {Kochan}\ \emph {et~al.}(2023)\citenamefont {Kochan},
  \citenamefont {Costa}, \citenamefont {Zhumagulov},\ and\ \citenamefont {{\v
  Z}uti{\'c}}}]{Kochan2023}%
  \BibitemOpen
  \bibfield  {author} {\bibinfo {author} {\bibfnamefont {D.}~\bibnamefont
  {Kochan}}, \bibinfo {author} {\bibfnamefont {A.}~\bibnamefont {Costa}},
  \bibinfo {author} {\bibfnamefont {I.}~\bibnamefont {Zhumagulov}}, \ and\
  \bibinfo {author} {\bibfnamefont {I.}~\bibnamefont {{\v Z}uti{\'c}}},\
  }\bibfield  {title} {\enquote {\bibinfo {title} {Phenomenological theory of
  the supercurrent diode effect: The {Lifshitz} invariant},}\ }\href@noop {}
  {\bibfield  {journal} {\bibinfo  {journal} {arXiv:2303.11975}\ } (\bibinfo
  {year} {2023})}\BibitemShut {NoStop}%
\bibitem [{\citenamefont {Agterberg}(2012)}]{Agterberg2012}%
  \BibitemOpen
  \bibfield  {author} {\bibinfo {author} {\bibfnamefont {D.~F.}\ \bibnamefont
  {Agterberg}},\ }\bibfield  {title} {\enquote {\bibinfo {title}
  {Magnetoelectric effects, helical phases, and {FFLO} phases},}\ }in\
  \href@noop {} {\emph {\bibinfo {booktitle} {Non-centrosymmetric
  Superconductors}}},\ \bibinfo {series} {Lecture Notes in Physics}, Vol.\
  \bibinfo {volume} {847},\ \bibinfo {editor} {edited by\ \bibinfo {editor}
  {\bibfnamefont {E.}~\bibnamefont {Bauer}}\ and\ \bibinfo {editor}
  {\bibfnamefont {M.}~\bibnamefont {Sigrist}}}\ (\bibinfo  {publisher}
  {Springer},\ \bibinfo {year} {2012})\ Chap.~\bibinfo {chapter} {5}, pp.\
  \bibinfo {pages} {155--170}\BibitemShut {NoStop}%
\bibitem [{\citenamefont {Xi}\ \emph {et~al.}(2016)\citenamefont {Xi},
  \citenamefont {Wang}, \citenamefont {Zhao}, \citenamefont {Park},
  \citenamefont {Law}, \citenamefont {Berger}, \citenamefont {Forr{\'o}},
  \citenamefont {Shan},\ and\ \citenamefont {Mak}}]{Xi2016}%
  \BibitemOpen
  \bibfield  {author} {\bibinfo {author} {\bibfnamefont {X.}~\bibnamefont
  {Xi}}, \bibinfo {author} {\bibfnamefont {Z.}~\bibnamefont {Wang}}, \bibinfo
  {author} {\bibfnamefont {W.}~\bibnamefont {Zhao}}, \bibinfo {author}
  {\bibfnamefont {J.-H.}\ \bibnamefont {Park}}, \bibinfo {author}
  {\bibfnamefont {K.~T.}\ \bibnamefont {Law}}, \bibinfo {author} {\bibfnamefont
  {H.}~\bibnamefont {Berger}}, \bibinfo {author} {\bibfnamefont
  {L.}~\bibnamefont {Forr{\'o}}}, \bibinfo {author} {\bibfnamefont
  {J.}~\bibnamefont {Shan}}, \ and\ \bibinfo {author} {\bibfnamefont {K.~F.}\
  \bibnamefont {Mak}},\ }\bibfield  {title} {\enquote {\bibinfo {title} {Ising
  pairing in superconducting {NbSe$_2$} atomic layers},}\ }\href@noop {}
  {\bibfield  {journal} {\bibinfo  {journal} {Nat. Phys.}\ }\textbf {\bibinfo
  {volume} {12}},\ \bibinfo {pages} {139} (\bibinfo {year} {2016})}\BibitemShut
  {NoStop}%
\bibitem [{\citenamefont {Costanzo}\ \emph {et~al.}(2016)\citenamefont
  {Costanzo}, \citenamefont {Jo}, \citenamefont {Berger},\ and\ \citenamefont
  {Morpurgo}}]{Costanzo2016}%
  \BibitemOpen
  \bibfield  {author} {\bibinfo {author} {\bibfnamefont {Davide}\ \bibnamefont
  {Costanzo}}, \bibinfo {author} {\bibfnamefont {Sanghyun}\ \bibnamefont {Jo}},
  \bibinfo {author} {\bibfnamefont {Helmuth}\ \bibnamefont {Berger}}, \ and\
  \bibinfo {author} {\bibfnamefont {Alberto~F.}\ \bibnamefont {Morpurgo}},\
  }\bibfield  {title} {\enquote {\bibinfo {title} {Gate-induced
  superconductivity in atomically thin {MoS$_2$} crystals},}\ }\href {\doibase
  10.1038/nnano.2015.314} {\bibfield  {journal} {\bibinfo  {journal} {Nature
  Nanotechnology}\ }\textbf {\bibinfo {volume} {11}},\ \bibinfo {pages}
  {339--344} (\bibinfo {year} {2016})}\BibitemShut {NoStop}%
\bibitem [{\citenamefont {Lu}\ \emph {et~al.}(2015)\citenamefont {Lu},
  \citenamefont {Zheliuk}, \citenamefont {Leermakers}, \citenamefont {Yuan},
  \citenamefont {Zeitler}, \citenamefont {Law},\ and\ \citenamefont
  {Ye}}]{Lu2015}%
  \BibitemOpen
  \bibfield  {author} {\bibinfo {author} {\bibfnamefont {J.~M.}\ \bibnamefont
  {Lu}}, \bibinfo {author} {\bibfnamefont {O.}~\bibnamefont {Zheliuk}},
  \bibinfo {author} {\bibfnamefont {I.}~\bibnamefont {Leermakers}}, \bibinfo
  {author} {\bibfnamefont {N.~F.~Q.}\ \bibnamefont {Yuan}}, \bibinfo {author}
  {\bibfnamefont {U.}~\bibnamefont {Zeitler}}, \bibinfo {author} {\bibfnamefont
  {K.~T.}\ \bibnamefont {Law}}, \ and\ \bibinfo {author} {\bibfnamefont
  {J.~T.}\ \bibnamefont {Ye}},\ }\bibfield  {title} {\enquote {\bibinfo {title}
  {Evidence for two-dimensional {Ising} superconductivity in gated
  {MoS$_2$}},}\ }\href@noop {} {\bibfield  {journal} {\bibinfo  {journal}
  {Science}\ }\textbf {\bibinfo {volume} {350}},\ \bibinfo {pages} {1353}
  (\bibinfo {year} {2015})}\BibitemShut {NoStop}%
\bibitem [{\citenamefont {Saito}\ \emph {et~al.}(2016)\citenamefont {Saito},
  \citenamefont {Nakamura}, \citenamefont {Bahramy}, \citenamefont {Kohama},
  \citenamefont {Ye}, \citenamefont {Kasahara}, \citenamefont {Nakagawa},
  \citenamefont {Onga}, \citenamefont {Tokunaga}, \citenamefont {Nojima},
  \citenamefont {Yanase},\ and\ \citenamefont {Iwasa}}]{Saito2016}%
  \BibitemOpen
  \bibfield  {author} {\bibinfo {author} {\bibfnamefont {Y.}~\bibnamefont
  {Saito}}, \bibinfo {author} {\bibfnamefont {Y.}~\bibnamefont {Nakamura}},
  \bibinfo {author} {\bibfnamefont {M.~S.}\ \bibnamefont {Bahramy}}, \bibinfo
  {author} {\bibfnamefont {Y.}~\bibnamefont {Kohama}}, \bibinfo {author}
  {\bibfnamefont {J.}~\bibnamefont {Ye}}, \bibinfo {author} {\bibfnamefont
  {Y.}~\bibnamefont {Kasahara}}, \bibinfo {author} {\bibfnamefont
  {Y.}~\bibnamefont {Nakagawa}}, \bibinfo {author} {\bibfnamefont
  {M.}~\bibnamefont {Onga}}, \bibinfo {author} {\bibfnamefont {M.}~\bibnamefont
  {Tokunaga}}, \bibinfo {author} {\bibfnamefont {T.}~\bibnamefont {Nojima}},
  \bibinfo {author} {\bibfnamefont {Y.}~\bibnamefont {Yanase}}, \ and\ \bibinfo
  {author} {\bibfnamefont {Y.}~\bibnamefont {Iwasa}},\ }\bibfield  {title}
  {\enquote {\bibinfo {title} {Superconductivity protected by spin-valley
  locking in ion-gated {MoS$_2$}},}\ }\href@noop {} {\bibfield  {journal}
  {\bibinfo  {journal} {Nat. Phys.}\ }\textbf {\bibinfo {volume} {12}},\
  \bibinfo {pages} {144} (\bibinfo {year} {2016})}\BibitemShut {NoStop}%
\bibitem [{\citenamefont {de~la Barrera}\ \emph {et~al.}(2018)\citenamefont
  {de~la Barrera}, \citenamefont {Sinko}, \citenamefont {Gopalan},
  \citenamefont {Sivadas}, \citenamefont {Seyler}, \citenamefont {Watanabe},
  \citenamefont {Taniguchi}, \citenamefont {Tsen}, \citenamefont {Xu},
  \citenamefont {Xiao},\ and\ \citenamefont {Hunt}}]{delaBarrera2018}%
  \BibitemOpen
  \bibfield  {author} {\bibinfo {author} {\bibfnamefont {S.~C.}\ \bibnamefont
  {de~la Barrera}}, \bibinfo {author} {\bibfnamefont {M.~R.}\ \bibnamefont
  {Sinko}}, \bibinfo {author} {\bibfnamefont {D.~P.}\ \bibnamefont {Gopalan}},
  \bibinfo {author} {\bibfnamefont {N.}~\bibnamefont {Sivadas}}, \bibinfo
  {author} {\bibfnamefont {K.~L.}\ \bibnamefont {Seyler}}, \bibinfo {author}
  {\bibfnamefont {K.}~\bibnamefont {Watanabe}}, \bibinfo {author}
  {\bibfnamefont {T.}~\bibnamefont {Taniguchi}}, \bibinfo {author}
  {\bibfnamefont {A.~W.}\ \bibnamefont {Tsen}}, \bibinfo {author}
  {\bibfnamefont {X.}~\bibnamefont {Xu}}, \bibinfo {author} {\bibfnamefont
  {D.}~\bibnamefont {Xiao}}, \ and\ \bibinfo {author} {\bibfnamefont {B.~M.}\
  \bibnamefont {Hunt}},\ }\bibfield  {title} {\enquote {\bibinfo {title}
  {Tuning {Ising} superconductivity with layer and spin-orbit coupling in
  two-dimensional transition-metal dichalcogenides},}\ }\href@noop {}
  {\bibfield  {journal} {\bibinfo  {journal} {Nat. Commun.}\ }\textbf {\bibinfo
  {volume} {9}},\ \bibinfo {pages} {1427} (\bibinfo {year} {2018})}\BibitemShut
  {NoStop}%
\bibitem [{\citenamefont {Sohn}\ \emph {et~al.}(2018)\citenamefont {Sohn},
  \citenamefont {Xi}, \citenamefont {He}, \citenamefont {Jiang}, \citenamefont
  {Wang}, \citenamefont {Kang}, \citenamefont {Park}, \citenamefont {Berger},
  \citenamefont {Forr{\'o}}, \citenamefont {Law}, \citenamefont {Shan},\ and\
  \citenamefont {Mak}}]{Sohn2018}%
  \BibitemOpen
  \bibfield  {author} {\bibinfo {author} {\bibfnamefont {E.}~\bibnamefont
  {Sohn}}, \bibinfo {author} {\bibfnamefont {X.}~\bibnamefont {Xi}}, \bibinfo
  {author} {\bibfnamefont {W.-Y.}\ \bibnamefont {He}}, \bibinfo {author}
  {\bibfnamefont {S.}~\bibnamefont {Jiang}}, \bibinfo {author} {\bibfnamefont
  {Z.}~\bibnamefont {Wang}}, \bibinfo {author} {\bibfnamefont {K.}~\bibnamefont
  {Kang}}, \bibinfo {author} {\bibfnamefont {J.-H.}\ \bibnamefont {Park}},
  \bibinfo {author} {\bibfnamefont {H.}~\bibnamefont {Berger}}, \bibinfo
  {author} {\bibfnamefont {L.}~\bibnamefont {Forr{\'o}}}, \bibinfo {author}
  {\bibfnamefont {K.~T.}\ \bibnamefont {Law}}, \bibinfo {author} {\bibfnamefont
  {J.}~\bibnamefont {Shan}}, \ and\ \bibinfo {author} {\bibfnamefont {K.~F.}\
  \bibnamefont {Mak}},\ }\bibfield  {title} {\enquote {\bibinfo {title} {An
  unusual continuous paramagnetic-limited superconducting phase transition in
  2d {NbSe$_2$}},}\ }\href@noop {} {\bibfield  {journal} {\bibinfo  {journal}
  {Nat. Mater.}\ }\textbf {\bibinfo {volume} {17}},\ \bibinfo {pages} {504}
  (\bibinfo {year} {2018})}\BibitemShut {NoStop}%
\bibitem [{\citenamefont {Dvir}\ \emph {et~al.}(2018)\citenamefont {Dvir},
  \citenamefont {Massee}, \citenamefont {Attias}, \citenamefont {Khodas},
  \citenamefont {Aprili}, \citenamefont {Quay},\ and\ \citenamefont
  {Steinberg}}]{Dvir2018}%
  \BibitemOpen
  \bibfield  {author} {\bibinfo {author} {\bibfnamefont {T.}~\bibnamefont
  {Dvir}}, \bibinfo {author} {\bibfnamefont {F.}~\bibnamefont {Massee}},
  \bibinfo {author} {\bibfnamefont {L.}~\bibnamefont {Attias}}, \bibinfo
  {author} {\bibfnamefont {M.}~\bibnamefont {Khodas}}, \bibinfo {author}
  {\bibfnamefont {M.}~\bibnamefont {Aprili}}, \bibinfo {author} {\bibfnamefont
  {C.~H.~L.}\ \bibnamefont {Quay}}, \ and\ \bibinfo {author} {\bibfnamefont
  {H.}~\bibnamefont {Steinberg}},\ }\bibfield  {title} {\enquote {\bibinfo
  {title} {Spectroscopy of bulk and few-layer superconducting {NbSe$_2$} with
  van der {Waals} tunnel junctions},}\ }\href@noop {} {\bibfield  {journal}
  {\bibinfo  {journal} {Nat. Commun.}\ }\textbf {\bibinfo {volume} {9}},\
  \bibinfo {pages} {598} (\bibinfo {year} {2018})}\BibitemShut {NoStop}%
\bibitem [{\citenamefont {Bulaevskii}\ \emph {et~al.}(1976)\citenamefont
  {Bulaevskii}, \citenamefont {Guseinov},\ and\ \citenamefont
  {Rusinov}}]{Bulaevskii1976}%
  \BibitemOpen
  \bibfield  {author} {\bibinfo {author} {\bibfnamefont {L.~N.}\ \bibnamefont
  {Bulaevskii}}, \bibinfo {author} {\bibfnamefont {A.~A.}\ \bibnamefont
  {Guseinov}}, \ and\ \bibinfo {author} {\bibfnamefont {A.~I.}\ \bibnamefont
  {Rusinov}},\ }\bibfield  {title} {\enquote {\bibinfo {title}
  {Superconductivity in crystals without symmetry centers},}\ }\href@noop {}
  {\bibfield  {journal} {\bibinfo  {journal} {Zh. Eksp. Teor. Fiz.}\ }\textbf
  {\bibinfo {volume} {71}},\ \bibinfo {pages} {2356} (\bibinfo {year}
  {1976})},\ \bibinfo {note} {[Sov. Phys. JETP {\bf 44}, 1243
  (1976)]}\BibitemShut {NoStop}%
\bibitem [{\citenamefont {Frigeri}\ \emph {et~al.}(2004)\citenamefont
  {Frigeri}, \citenamefont {Agterberg}, \citenamefont {Koga},\ and\
  \citenamefont {Sigrist}}]{Frigeri2004}%
  \BibitemOpen
  \bibfield  {author} {\bibinfo {author} {\bibfnamefont {P.~A.}\ \bibnamefont
  {Frigeri}}, \bibinfo {author} {\bibfnamefont {D.~F.}\ \bibnamefont
  {Agterberg}}, \bibinfo {author} {\bibfnamefont {A.}~\bibnamefont {Koga}}, \
  and\ \bibinfo {author} {\bibfnamefont {M.}~\bibnamefont {Sigrist}},\
  }\bibfield  {title} {\enquote {\bibinfo {title} {Superconductivity without
  inversion symmetry: {MnSi} versus {CePt$_3$Si}},}\ }\href@noop {} {\bibfield
  {journal} {\bibinfo  {journal} {Phys. Rev. Lett.}\ }\textbf {\bibinfo
  {volume} {92}},\ \bibinfo {pages} {097001} (\bibinfo {year}
  {2004})}\BibitemShut {NoStop}%
\bibitem [{\citenamefont {Ili{\'c}}\ \emph {et~al.}(2017)\citenamefont
  {Ili{\'c}}, \citenamefont {Meyer},\ and\ \citenamefont {Houzet}}]{Ilic2017}%
  \BibitemOpen
  \bibfield  {author} {\bibinfo {author} {\bibfnamefont {S.}~\bibnamefont
  {Ili{\'c}}}, \bibinfo {author} {\bibfnamefont {J.~S.}\ \bibnamefont {Meyer}},
  \ and\ \bibinfo {author} {\bibfnamefont {M.}~\bibnamefont {Houzet}},\
  }\bibfield  {title} {\enquote {\bibinfo {title} {Enhancement of the upper
  critical field in disordered transition metal dichalcogenide monolayers},}\
  }\href@noop {} {\bibfield  {journal} {\bibinfo  {journal} {Phys. Rev. Lett.}\
  }\textbf {\bibinfo {volume} {119}},\ \bibinfo {pages} {117001} (\bibinfo
  {year} {2017})}\BibitemShut {NoStop}%
\bibitem [{\citenamefont {Sosenko}\ \emph {et~al.}(2017)\citenamefont
  {Sosenko}, \citenamefont {Zhang},\ and\ \citenamefont {Aji}}]{Sosenko2017}%
  \BibitemOpen
  \bibfield  {author} {\bibinfo {author} {\bibfnamefont {E.}~\bibnamefont
  {Sosenko}}, \bibinfo {author} {\bibfnamefont {J.}~\bibnamefont {Zhang}}, \
  and\ \bibinfo {author} {\bibfnamefont {V.}~\bibnamefont {Aji}},\ }\bibfield
  {title} {\enquote {\bibinfo {title} {Unconventional superconductivity and
  anomalous response in hole-doped transition metal dichalcogenides},}\
  }\href@noop {} {\bibfield  {journal} {\bibinfo  {journal} {Phys. Rev. B}\
  }\textbf {\bibinfo {volume} {95}},\ \bibinfo {pages} {144508} (\bibinfo
  {year} {2017})}\BibitemShut {NoStop}%
\bibitem [{\citenamefont {M{\"o}ckli}\ and\ \citenamefont
  {Khodas}(2018)}]{Mockli2018}%
  \BibitemOpen
  \bibfield  {author} {\bibinfo {author} {\bibfnamefont {D.}~\bibnamefont
  {M{\"o}ckli}}\ and\ \bibinfo {author} {\bibfnamefont {M.}~\bibnamefont
  {Khodas}},\ }\bibfield  {title} {\enquote {\bibinfo {title} {Robust
  parity-mixed superconductivity in disordered monolayer transition metal
  dichalcogenides},}\ }\href@noop {} {\bibfield  {journal} {\bibinfo  {journal}
  {Phys. Rev. B}\ }\textbf {\bibinfo {volume} {98}},\ \bibinfo {pages} {144518}
  (\bibinfo {year} {2018})}\BibitemShut {NoStop}%
\bibitem [{\citenamefont {M{\"o}ckli}\ and\ \citenamefont
  {Khodas}(2019)}]{Mockli2019}%
  \BibitemOpen
  \bibfield  {author} {\bibinfo {author} {\bibfnamefont {D.}~\bibnamefont
  {M{\"o}ckli}}\ and\ \bibinfo {author} {\bibfnamefont {M.}~\bibnamefont
  {Khodas}},\ }\bibfield  {title} {\enquote {\bibinfo {title} {Magnetic-field
  induced $s+if$ pairing in {Ising} superconductors},}\ }\href@noop {}
  {\bibfield  {journal} {\bibinfo  {journal} {Phys. Rev. B}\ }\textbf {\bibinfo
  {volume} {99}},\ \bibinfo {pages} {180505(R)} (\bibinfo {year}
  {2019})}\BibitemShut {NoStop}%
\bibitem [{\citenamefont {M{\"o}ckli}\ and\ \citenamefont
  {Khodas}(2020)}]{Mockli2020}%
  \BibitemOpen
  \bibfield  {author} {\bibinfo {author} {\bibfnamefont {D.}~\bibnamefont
  {M{\"o}ckli}}\ and\ \bibinfo {author} {\bibfnamefont {M.}~\bibnamefont
  {Khodas}},\ }\bibfield  {title} {\enquote {\bibinfo {title} {Ising
  superconductors: Interplay of magnetic field, triplet channels, and
  disorder},}\ }\href@noop {} {\bibfield  {journal} {\bibinfo  {journal} {Phys.
  Rev. B}\ }\textbf {\bibinfo {volume} {101}},\ \bibinfo {pages} {014510}
  (\bibinfo {year} {2020})}\BibitemShut {NoStop}%
\bibitem [{\citenamefont {Haim}\ \emph {et~al.}(2020)\citenamefont {Haim},
  \citenamefont {M{\"o}ckli},\ and\ \citenamefont {Khodas}}]{Haim2020}%
  \BibitemOpen
  \bibfield  {author} {\bibinfo {author} {\bibfnamefont {M.}~\bibnamefont
  {Haim}}, \bibinfo {author} {\bibfnamefont {D.}~\bibnamefont {M{\"o}ckli}}, \
  and\ \bibinfo {author} {\bibfnamefont {M.}~\bibnamefont {Khodas}},\
  }\bibfield  {title} {\enquote {\bibinfo {title} {Signatures of triplet
  correlations in density of states of {Ising} superconductors},}\ }\href@noop
  {} {\bibfield  {journal} {\bibinfo  {journal} {Phys. Rev. B}\ }\textbf
  {\bibinfo {volume} {102}},\ \bibinfo {pages} {214513} (\bibinfo {year}
  {2020})}\BibitemShut {NoStop}%
\bibitem [{\citenamefont {Tang}\ \emph {et~al.}(2021)\citenamefont {Tang},
  \citenamefont {Bruder},\ and\ \citenamefont {Belzig}}]{Tang2021}%
  \BibitemOpen
  \bibfield  {author} {\bibinfo {author} {\bibfnamefont {G.}~\bibnamefont
  {Tang}}, \bibinfo {author} {\bibfnamefont {C.}~\bibnamefont {Bruder}}, \ and\
  \bibinfo {author} {\bibfnamefont {W.}~\bibnamefont {Belzig}},\ }\bibfield
  {title} {\enquote {\bibinfo {title} {Magnetic field-induced ``mirage'' gap in
  an {Ising} superconductor},}\ }\href@noop {} {\bibfield  {journal} {\bibinfo
  {journal} {Phys. Rev. Lett.}\ }\textbf {\bibinfo {volume} {126}},\ \bibinfo
  {pages} {237001} (\bibinfo {year} {2021})}\BibitemShut {NoStop}%
\bibitem [{\citenamefont {Bychkov}\ and\ \citenamefont
  {Rashba}(1984)}]{BychkovRashba1984}%
  \BibitemOpen
  \bibfield  {author} {\bibinfo {author} {\bibfnamefont {Y.~A.}\ \bibnamefont
  {Bychkov}}\ and\ \bibinfo {author} {\bibfnamefont {E.~I.}\ \bibnamefont
  {Rashba}},\ }\bibfield  {title} {\enquote {\bibinfo {title} {Properties of a
  {2D} electron gas with lifted spectral degeneracy},}\ }\href@noop {}
  {\bibfield  {journal} {\bibinfo  {journal} {Pis'ma Zh. Eksp. Teor. Fiz.}\
  }\textbf {\bibinfo {volume} {39}},\ \bibinfo {pages} {66} (\bibinfo {year}
  {1984})},\ \bibinfo {note} {[JETP Lett. {\bf 39}, 78 (1984)]}\BibitemShut
  {NoStop}%
\bibitem [{\citenamefont {Gor'kov}\ and\ \citenamefont
  {Rashba}(2001)}]{GorkovRashba2001}%
  \BibitemOpen
  \bibfield  {author} {\bibinfo {author} {\bibfnamefont {L.~P.}\ \bibnamefont
  {Gor'kov}}\ and\ \bibinfo {author} {\bibfnamefont {E.~I.}\ \bibnamefont
  {Rashba}},\ }\bibfield  {title} {\enquote {\bibinfo {title} {Superconducting
  2d system with lifted spin degeneracy: Mixed singlet-triplet state},}\
  }\href@noop {} {\bibfield  {journal} {\bibinfo  {journal} {Phys. Rev. Lett.}\
  }\textbf {\bibinfo {volume} {87}},\ \bibinfo {pages} {037004} (\bibinfo
  {year} {2001})}\BibitemShut {NoStop}%
\bibitem [{\citenamefont {Harms}\ \emph {et~al.}(2026)\citenamefont {Harms},
  \citenamefont {Hein},\ and\ \citenamefont {Belzig}}]{Harms2026}%
  \BibitemOpen
  \bibfield  {author} {\bibinfo {author} {\bibfnamefont {J.~S.}\ \bibnamefont
  {Harms}}, \bibinfo {author} {\bibfnamefont {M.}~\bibnamefont {Hein}}, \ and\
  \bibinfo {author} {\bibfnamefont {W.}~\bibnamefont {Belzig}},\ }\href
  {https://arxiv.org/abs/2512.01910} {\enquote {\bibinfo {title} {Collapse of
  the superconducting order parameter in {Ising} superconductors with {Rashba}
  spin-orbit coupling},}\ } (\bibinfo {year} {2026}),\ \Eprint
  {http://arxiv.org/abs/2512.01910} {arXiv:2512.01910 [cond-mat.supr-con]}
  \BibitemShut {NoStop}%
\bibitem [{\citenamefont {Hani\ifmmode~\check{s}\else \v{s}\fi{}}\ \emph
  {et~al.}(2024)\citenamefont {Hani\ifmmode~\check{s}\else \v{s}\fi{}},
  \citenamefont {Milivojevi\ifmmode~\acute{c}\else \'{c}\fi{}},\ and\
  \citenamefont {Gmitra}}]{Hanis2024}%
  \BibitemOpen
  \bibfield  {author} {\bibinfo {author} {\bibfnamefont {Jozef}\ \bibnamefont
  {Hani\ifmmode~\check{s}\else \v{s}\fi{}}}, \bibinfo {author} {\bibfnamefont
  {Marko}\ \bibnamefont {Milivojevi\ifmmode~\acute{c}\else \'{c}\fi{}}}, \ and\
  \bibinfo {author} {\bibfnamefont {Martin}\ \bibnamefont {Gmitra}},\
  }\bibfield  {title} {\enquote {\bibinfo {title} {Distinguishing nodal and
  nonunitary superconductivity in quasiparticle interference of an {Ising}
  superconductor with rashba spin-orbit coupling: The example of
  {NbSe$_{2}$}},}\ }\href {\doibase 10.1103/PhysRevB.110.104502} {\bibfield
  {journal} {\bibinfo  {journal} {Phys. Rev. B}\ }\textbf {\bibinfo {volume}
  {110}},\ \bibinfo {pages} {104502} (\bibinfo {year} {2024})}\BibitemShut
  {NoStop}%
\bibitem [{\citenamefont {Bauer}\ and\ \citenamefont
  {Sigrist}(2012)}]{BauerSigrist2012}%
  \BibitemOpen
  \bibfield  {author} {\bibinfo {author} {\bibfnamefont {E.}~\bibnamefont
  {Bauer}}\ and\ \bibinfo {author} {\bibfnamefont {M.}~\bibnamefont
  {Sigrist}},\ }\href@noop {} {\emph {\bibinfo {title} {Non-Centrosymmetric
  Superconductors: Introduction and Overview}}},\ \bibinfo {series} {Lecture
  Notes in Physics}, Vol.\ \bibinfo {volume} {847}\ (\bibinfo  {publisher}
  {Springer},\ \bibinfo {address} {Berlin, Heidelberg},\ \bibinfo {year}
  {2012})\BibitemShut {NoStop}%
\bibitem [{\citenamefont {Smidman}\ \emph {et~al.}(2017)\citenamefont
  {Smidman}, \citenamefont {Salamon}, \citenamefont {Yuan},\ and\ \citenamefont
  {Agterberg}}]{Smidman2017}%
  \BibitemOpen
  \bibfield  {author} {\bibinfo {author} {\bibfnamefont {M.}~\bibnamefont
  {Smidman}}, \bibinfo {author} {\bibfnamefont {M.~B.}\ \bibnamefont
  {Salamon}}, \bibinfo {author} {\bibfnamefont {H.~Q.}\ \bibnamefont {Yuan}}, \
  and\ \bibinfo {author} {\bibfnamefont {D.~F.}\ \bibnamefont {Agterberg}},\
  }\bibfield  {title} {\enquote {\bibinfo {title} {Superconductivity and
  spin-orbit coupling in non-centrosymmetric materials: a review},}\
  }\href@noop {} {\bibfield  {journal} {\bibinfo  {journal} {Rep. Prog. Phys.}\
  }\textbf {\bibinfo {volume} {80}},\ \bibinfo {pages} {036501} (\bibinfo
  {year} {2017})}\BibitemShut {NoStop}%
\bibitem [{\citenamefont {Kuzmanovi\ifmmode~\acute{c}\else \'{c}\fi{}}\ \emph
  {et~al.}(2022)\citenamefont {Kuzmanovi\ifmmode~\acute{c}\else \'{c}\fi{}},
  \citenamefont {Dvir}, \citenamefont {LeBoeuf}, \citenamefont
  {Ili\ifmmode~\acute{c}\else \'{c}\fi{}}, \citenamefont {Haim}, \citenamefont
  {M\"ockli}, \citenamefont {Kramer}, \citenamefont {Khodas}, \citenamefont
  {Houzet}, \citenamefont {Meyer}, \citenamefont {Aprili}, \citenamefont
  {Steinberg},\ and\ \citenamefont {Quay}}]{Kuzmanovic2022}%
  \BibitemOpen
  \bibfield  {author} {\bibinfo {author} {\bibfnamefont {M.}~\bibnamefont
  {Kuzmanovi\ifmmode~\acute{c}\else \'{c}\fi{}}}, \bibinfo {author}
  {\bibfnamefont {T.}~\bibnamefont {Dvir}}, \bibinfo {author} {\bibfnamefont
  {D.}~\bibnamefont {LeBoeuf}}, \bibinfo {author} {\bibfnamefont
  {S.}~\bibnamefont {Ili\ifmmode~\acute{c}\else \'{c}\fi{}}}, \bibinfo {author}
  {\bibfnamefont {M.}~\bibnamefont {Haim}}, \bibinfo {author} {\bibfnamefont
  {D.}~\bibnamefont {M\"ockli}}, \bibinfo {author} {\bibfnamefont
  {S.}~\bibnamefont {Kramer}}, \bibinfo {author} {\bibfnamefont
  {M.}~\bibnamefont {Khodas}}, \bibinfo {author} {\bibfnamefont
  {M.}~\bibnamefont {Houzet}}, \bibinfo {author} {\bibfnamefont {J.~S.}\
  \bibnamefont {Meyer}}, \bibinfo {author} {\bibfnamefont {M.}~\bibnamefont
  {Aprili}}, \bibinfo {author} {\bibfnamefont {H.}~\bibnamefont {Steinberg}}, \
  and\ \bibinfo {author} {\bibfnamefont {C.~H.~L.}\ \bibnamefont {Quay}},\
  }\bibfield  {title} {\enquote {\bibinfo {title} {Tunneling spectroscopy of
  few-monolayer {${\mathrm{NbSe}}_{2}$} in high magnetic fields: Triplet
  superconductivity and {Ising} protection},}\ }\href {\doibase
  10.1103/PhysRevB.106.184514} {\bibfield  {journal} {\bibinfo  {journal}
  {Phys. Rev. B}\ }\textbf {\bibinfo {volume} {106}},\ \bibinfo {pages}
  {184514} (\bibinfo {year} {2022})}\BibitemShut {NoStop}%
\bibitem [{\citenamefont {Das}\ \emph {et~al.}(2023)\citenamefont {Das},
  \citenamefont {Paudyal}, \citenamefont {Margine}, \citenamefont {Agterberg},\
  and\ \citenamefont {Mazin}}]{Das2023}%
  \BibitemOpen
  \bibfield  {author} {\bibinfo {author} {\bibfnamefont {S.}~\bibnamefont
  {Das}}, \bibinfo {author} {\bibfnamefont {H.}~\bibnamefont {Paudyal}},
  \bibinfo {author} {\bibfnamefont {E.~R.}\ \bibnamefont {Margine}}, \bibinfo
  {author} {\bibfnamefont {D.~F.}\ \bibnamefont {Agterberg}}, \ and\ \bibinfo
  {author} {\bibfnamefont {I.~I.}\ \bibnamefont {Mazin}},\ }\bibfield  {title}
  {\enquote {\bibinfo {title} {Electron-phonon coupling and spin fluctuations
  in the {Ising} superconductor {NbSe$_2$}},}\ }\href {\doibase
  10.1038/s41524-023-01017-4} {\bibfield  {journal} {\bibinfo  {journal} {npj
  Computational Materials}\ }\textbf {\bibinfo {volume} {9}},\ \bibinfo {pages}
  {66} (\bibinfo {year} {2023})}\BibitemShut {NoStop}%
\bibitem [{\citenamefont {Roy}\ \emph {et~al.}(2024)\citenamefont {Roy},
  \citenamefont {Kreisel}, \citenamefont {Andersen},\ and\ \citenamefont
  {Mukherjee}}]{Roy2024}%
  \BibitemOpen
  \bibfield  {author} {\bibinfo {author} {\bibfnamefont {Subhojit}\
  \bibnamefont {Roy}}, \bibinfo {author} {\bibfnamefont {Andreas}\ \bibnamefont
  {Kreisel}}, \bibinfo {author} {\bibfnamefont {Brian~M}\ \bibnamefont
  {Andersen}}, \ and\ \bibinfo {author} {\bibfnamefont {Shantanu}\ \bibnamefont
  {Mukherjee}},\ }\bibfield  {title} {\enquote {\bibinfo {title}
  {Unconventional pairing in {Ising} superconductors: application to monolayer
  {NbSe$_2$}},}\ }\href {\doibase 10.1088/2053-1583/ad7b53} {\bibfield
  {journal} {\bibinfo  {journal} {2D Materials}\ }\textbf {\bibinfo {volume}
  {12}},\ \bibinfo {pages} {015004} (\bibinfo {year} {2024})}\BibitemShut
  {NoStop}%
\bibitem [{\citenamefont {Roy}\ \emph {et~al.}(2026)\citenamefont {Roy},
  \citenamefont {Kreisel}, \citenamefont {Andersen},\ and\ \citenamefont
  {Mukherjee}}]{Roy2026}%
  \BibitemOpen
  \bibfield  {author} {\bibinfo {author} {\bibfnamefont {Subhojit}\
  \bibnamefont {Roy}}, \bibinfo {author} {\bibfnamefont {Andreas}\ \bibnamefont
  {Kreisel}}, \bibinfo {author} {\bibfnamefont {Brian~M.}\ \bibnamefont
  {Andersen}}, \ and\ \bibinfo {author} {\bibfnamefont {Shantanu}\ \bibnamefont
  {Mukherjee}},\ }\bibfield  {title} {\enquote {\bibinfo {title}
  {Unconventional superconductivity in monolayer transition metal
  dichalcogenides},}\ }\href {\doibase 10.1103/6rc6-mm3b} {\bibfield  {journal}
  {\bibinfo  {journal} {Phys. Rev. B}\ }\textbf {\bibinfo {volume} {113}},\
  \bibinfo {pages} {014506} (\bibinfo {year} {2026})}\BibitemShut {NoStop}%
\bibitem [{\citenamefont {Horhold}\ \emph {et~al.}(2023)\citenamefont
  {Horhold}, \citenamefont {Graf}, \citenamefont {Marganska},\ and\
  \citenamefont {Grifoni}}]{Horhold2023}%
  \BibitemOpen
  \bibfield  {author} {\bibinfo {author} {\bibfnamefont {Sebastian}\
  \bibnamefont {Horhold}}, \bibinfo {author} {\bibfnamefont {Juliane}\
  \bibnamefont {Graf}}, \bibinfo {author} {\bibfnamefont {Magdalena}\
  \bibnamefont {Marganska}}, \ and\ \bibinfo {author} {\bibfnamefont {Milena}\
  \bibnamefont {Grifoni}},\ }\bibfield  {title} {\enquote {\bibinfo {title}
  {Two-bands {Ising} superconductivity from {Coulomb} interactions in monolayer
  {NbSe$_2$}},}\ }\href {\doibase 10.1088/2053-1583/acb21d} {\bibfield
  {journal} {\bibinfo  {journal} {2D Materials}\ }\textbf {\bibinfo {volume}
  {10}},\ \bibinfo {pages} {025008} (\bibinfo {year} {2023})}\BibitemShut
  {NoStop}%
\bibitem [{\citenamefont {Sohier}\ \emph {et~al.}(2025)\citenamefont {Sohier},
  \citenamefont {Gibertini}, \citenamefont {Martin},\ and\ \citenamefont
  {Morpurgo}}]{Sohier2025}%
  \BibitemOpen
  \bibfield  {author} {\bibinfo {author} {\bibfnamefont {Thibault}\
  \bibnamefont {Sohier}}, \bibinfo {author} {\bibfnamefont {Marco}\
  \bibnamefont {Gibertini}}, \bibinfo {author} {\bibfnamefont {Ivar}\
  \bibnamefont {Martin}}, \ and\ \bibinfo {author} {\bibfnamefont {Alberto~F.}\
  \bibnamefont {Morpurgo}},\ }\bibfield  {title} {\enquote {\bibinfo {title}
  {Unconventional gate-induced superconductivity in transition-metal
  dichalcogenides},}\ }\href {\doibase 10.1103/PhysRevResearch.7.013290}
  {\bibfield  {journal} {\bibinfo  {journal} {Phys. Rev. Res.}\ }\textbf
  {\bibinfo {volume} {7}},\ \bibinfo {pages} {013290} (\bibinfo {year}
  {2025})}\BibitemShut {NoStop}%
\bibitem [{\citenamefont {Fulde}\ and\ \citenamefont
  {Ferrell}(1964)}]{FuldeFerrell1964}%
  \BibitemOpen
  \bibfield  {author} {\bibinfo {author} {\bibfnamefont {P.}~\bibnamefont
  {Fulde}}\ and\ \bibinfo {author} {\bibfnamefont {R.~A.}\ \bibnamefont
  {Ferrell}},\ }\bibfield  {title} {\enquote {\bibinfo {title}
  {Superconductivity in a strong spin-exchange field},}\ }\href@noop {}
  {\bibfield  {journal} {\bibinfo  {journal} {Phys. Rev.}\ }\textbf {\bibinfo
  {volume} {135}},\ \bibinfo {pages} {A550} (\bibinfo {year}
  {1964})}\BibitemShut {NoStop}%
\bibitem [{\citenamefont {Larkin}\ and\ \citenamefont
  {Ovchinnikov}(1964)}]{LarkinOvchinnikov1964}%
  \BibitemOpen
  \bibfield  {author} {\bibinfo {author} {\bibfnamefont {A.~I.}\ \bibnamefont
  {Larkin}}\ and\ \bibinfo {author} {\bibfnamefont {Y.~N.}\ \bibnamefont
  {Ovchinnikov}},\ }\bibfield  {title} {\enquote {\bibinfo {title} {Nonuniform
  state of superconductors},}\ }\href@noop {} {\bibfield  {journal} {\bibinfo
  {journal} {Zh. Eksp. Teor. Fiz.}\ }\textbf {\bibinfo {volume} {47}},\
  \bibinfo {pages} {1136} (\bibinfo {year} {1964})},\ \bibinfo {note} {[Sov.
  Phys. JETP {\bf 20}, 762 (1965)]}\BibitemShut {NoStop}%
\bibitem [{\citenamefont {{Anthropic}}(2026)}]{Claude2026}%
  \BibitemOpen
  \bibfield  {author} {\bibinfo {author} {\bibnamefont {{Anthropic}}},\
  }\href@noop {} {\enquote {\bibinfo {title} {Claude [large language model]},}\
  }\bibinfo {howpublished} {\url{https://claude.ai}} (\bibinfo {year} {2026}),\
  \bibinfo {note} {version: Claude Fable 5; used June--July 2026}\BibitemShut
  {NoStop}%
\end{thebibliography}%

\end{document}